\documentclass [] {aa} 
\usepackage{epsfig, graphicx,longtable,lscape}
\usepackage{amsmath}
\usepackage{amssymb}
\usepackage{lscape}

\begin{document}

%\thesaurus{06(08.01.1; 08.03.02; 08.16.2)}

%\titlerunning{} 
\title{Abundances of refractory elements in the atmospheres of stars with extrasolar planets\thanks{Based on
    observations collected at the La Silla Observatory, ESO (Chile),
    with {\footnotesize CORALIE} spectrograph at the 1.2 m
    Euler Swiss telescope, and with the {\footnotesize FEROS} spectrograph    
    at the 1.52 m ESO telescope, at the Paranal Observatory, ESO (Chile), using the UVES spectrograph at the VLT/UT2 Kueyen
    telescope, and with the UES and SARG spectrographs at the 4 m William Hershel Telescope (WHT) and the 3.5 m TNG,
    respectively, both at La Palma (Canary Islands).}}

%\subtitle{}

\author{G.~Gilli\inst{1,2} \and G.~Israelian\inst{2} \and A.~Ecuvillon\inst{2} \and  N.C.~Santos\inst{3,4}  \and M.~Mayor\inst{4}} 

\offprints{G.~Israelian, \email{gil@iac.es}}

\institute{Dipartimento di Astronomia, Universit\`a degli Studi di Padova, Italy \and Instituto de Astrof\'isica de Canarias, 
E--38200 La Laguna, Tenerife, Spain \and Centro de Astronomia e Astrof{\'\i}sica da Universidade 
de Lisboa, Observat\'orio Astron\'omico de Lisboa, Tapada da Ajuda,
1349-018 Lisboa, Portugal \and Observatoire de Gen\`eve, 51 ch.  des
	   Maillettes, CH--1290 Sauverny, Switzerland}

\date{Received 19 Jul 2005/ Accepted 06 Dec 2005 } 

\titlerunning{Abundances of refractory elements} 

%--------------------------------------------------------------------------

\abstract
   {}
   {This work presents a uniform and homogeneous study of chemical abundances of refractory elements in 101 stars with and 94 without known planetary companions. We carry out an in-depth investigation of the abundances of Si, Ca, Sc, Ti, V, Cr, Mn, Co, Ni, Na, Mg and Al. The new comparison sample, spanning the metallicity range $-0.70< [$Fe$/$H$]< 0.50$, fills the gap that previously existed, mainly at high metallicities, in the number of stars without known planets.}
   {We used an enlarged set of data including new observations, especially for the field ``single'' comparison stars . The line list
    previously studied by other authors was improved: on average we analysed 90 spectral lines in every spectrum and carefully measured more than 16\,600 equivalent 
widths (EW) to calculate the abundances.}
   {We investigate possible differences between the chemical abundances of the two groups of stars, both with and without planets. The 
results are globally comparable to those obtained by other authors, and in most cases the abundance trends of planet-host stars are
very similar to those of the comparison sample. }
   {This work represents a step towards the comprehension of recently discovered planetary
systems. These results could also be useful for verifying galactic models at high metallicities and consequently improve our knowledge of
stellar nucleosynthesis and galactic chemical evolution.}

   \keywords{Stars: abundances -- Stars: planetary systems -- 
Galaxy: solar neighbourhood -- Galaxy: evolution}

\maketitle

\section{Introduction}
Over the last ten years a large  number of stars harbouring planets have been found. The first giant planet was discovered 
 around 51 Peg by Mayor \& Queloz (1995) and there are more than 150 planetary-mass companions presently known orbiting solar-type stars.
The growing number of extrasolar planets\footnote{A complete updated table of known planets can be found
at https://obswww.unige.ch/exoplanets.} has activated intensive study of these objects and their parent
stars, and nowadays extensive studies of the properties of the new planetary systems are conceivable. Several spectroscopic analyses 
of iron
abundances (Gonzalez 1998; Gonzalez et al.~2001; Laws et al.~2003; Santos et al.~2001a,2001b, 2003, 2004a,b, 2005) have suggested that 
planet-host stars are more metal-rich than field dwarfs.
These results show that the probability of finding a planet is a strongly increasing function of  stellar metallicity, at least for
[Fe/H] above solar value. 
Two interpretations, the \textit{self-enrichment hypothesis} and the \textit{primordial hypothesis}, have  been proposed  to explain a possible connection  between the metallicity excess and the
presence of planets. The former  considers that the 
observed iron overabundances derives from the accretion of a large amount of rocky planetesimal material on to the star (Gonzalez 1997).
The latter, the ``primordial scenario'', suggests that the iron excess in stars with planets just reflects the high metal content 
of the
protoplanetary cloud from which stars and planets were formed (Santos et al.~2000, 2001a,2001b).

In this context, abundance trends of chemical species other than iron can give important clues in this debate, so that discriminating
between these two possibilities will help in understanding how planetary systems form. Efforts have been made to analyse the
chemical abundances of light elements (e.g.\ Garc\'ia L\'opez \& P\'erez de Taoro 1998; Gonzalez \& Laws 2000; Ryan 2000; Deliyannis 
et al.~ 2000; Israelian et al.~ 2003a, 2004; Santos et al.~ 2002, 2004b), as well as other metals (e.g.\ Gonzalez \& Laws 2000; 
Gonzalez
et al.~2001; Smith et al.~2001; Takeda et al.~2001; Sadakane et al.~2002; Fisher \& Valenti 2005). Most of these authors considered  
inhomogeneous comparison
samples of field dwarfs from the literature that might be a source of systematic errors. 
 The majority of these studies  support the primordial hypothesis (Pinsonneault et al.~2001; 
 Santos et al.~2001a,2001b, 2002; Sadakane et al.~2002),
  but evidence of pollution has been
 found for a few cases (Israelian et al.~2001,2003a,2003b; Low \& Gonzalez 2001).

Unbiased and homogeneous studies of Fe abundance in stars with and without planets have been performed by Santos et al.\ (2001a, 2003, 
2004a, 2005). 
Similar studies have been recently carried out for elements other than iron. Refractory elements (some $\alpha$ and Fe-group elements)
have been analysed by Bodaghee et al.~(2003); all volatile elements (C, S, Zn, N, O) have been studied by Ecuvillon et al.~(2004a,2004b, 
2005) and Na, Al and Mg by Beir\~ao et al.~(2005).
These studies all required a uniform high-metallicity comparison, given the lack of ``single'' field stars in the data for
[Fe/H]$>0.1$. We provide results using new and more precise atmospheric parameters from high
quality spectra and also complete the high-metallicity comparison between the two samples.

In this paper we present a detailed, homogeneous and uniform study of Si, Ca, Ti, Sc,
V, Cr, Mn, Co, Ni, Na, Mg and Al in a set of 101 planet-harbouring stars and a group of 94 solar-type stars with no known
planets in the metallicity range $-0.70< $[Fe/H] $<0.45$. We have made use of about 16\,600 EWs measured to calculate the [$X$/H] ratio for 
each element and have plotted the results in the [$X$/$H$]
vs.\ [Fe/H] and [X/Fe] vs.\ [Fe/H] planes.

\begin{table}[!h]
\centering
\caption{ \small{Atomic parameters of the spectral lines used for every element. Col.~1: wavelength (in \AA); Col.~2: excitation
energy of the lower energy level in the transition (in eV); Col.~3: oscillator strengths based on an inverse solar analysis.}}							
 \scriptsize
 \label{linee}
   	\begin{tabular}{ccrcccr}	
	\hline\hline
   	\noalign{\smallskip}
   	\noalign{\smallskip}
	$\lambda$ & $\chi_{l}$ & $\log{gf}$ & &  $\lambda$ & $\chi_{l}$ & $\log{gf}$\\
	\noalign{\smallskip}
	\noalign{\smallskip}
	\hline
	\noalign{\smallskip}
	\multicolumn{3}{l}{\textbf{Si} I ; $\log \epsilon_{\circ} = 7.55$  $A =14$} &
	 & \multicolumn{3}{l}{\textbf{Cr} I ; $\log \epsilon_{\circ} = 5.67$  $A = 24$}  \\
	5665.56 & 4.92 & $-$1.980 & &    5304.18 & 3.46 & $-$0.680 \\
	5690.43 & 4.93 & $-$1.790 & & 	 5312.86 & 3.45 & $-$0.580  \\
	5701.10 & 4.93 & $-$2.020 & & 	 5318.77 & 3.44 & $-$0.710 \\
	5772.14 & 5.08 & $-$1.620 & & 	 5480.51 & 3.50 & $-$0.830 \\
	5793.09 & 4.93 & $-$1.910 & &    5574.39 & 4.45 & $-$0.480 \\
	5948.55 & 5.08 & $-$1.110 & & 	 5783.07 & 3.32 & $-$0.400 \\
	6125.02 & 5.61 & $-$1.520 & & 	 5783.87 & 3.32 & $-$0.150 \\
	6142.49 & 5.62 & $-$1.480 & &    5787.92 & 3.32 & $-$0.110 \\
	6145.02 & 5.61 & $-$1.400 & & 	 \multicolumn{3}{l}{\textbf{Mn} I ; $\log \epsilon_{\circ} = 5.39$ $A= 25$}\\
	6155.15 & 5.62 & $-$0.750 & & 	 4265.92 & 2.94 & $-$ 0.440 \\  		  
	6721.86 & 5.86 & $-$1.090 & & 	 4470.13 & 2.94 & $-$ 0.550 \\
	\multicolumn{3}{l}{\textbf{Ca} I ; $\log \epsilon_{\circ} = 6.36$ $A =$ 20}& & 4502.13 & 2.92 & $-$ 0.490 \\
	5512.98 & 2.93 & $-$0.440 & &     5399.47 & 3.85 & $-$0.0969 \\
	5581.97 & 2.52 & $-$0.650 & &     5413.68 & 3.85 & $-$0.470 \\
	5590.12 & 2.52 & $-$0.710 & &     5432.54 & 0.00 & $-$3.620  \\
	5867.56 & 2.93 & $-$1.590 & &     6440.93 & 3.77 & $-$1.250 \\  		    
	6161.29 & 2.52 & $-$1.220 & & 	  \multicolumn{3}{l}{\textbf{Co} I ; $\log \epsilon_{\circ} = 4.92$ $A = 27$} \\
	6166.44 & 2.52 & $-$1.120 & & 	   5301.04 & 1.71 & $-$1.930 \\ 			     
	6169.05 & 2.52 & $-$0.730 & & 	  5325.27 & 4.02 & $-$0.120  \\
	6169.56 & 2.52 & $-$0.440 & & 	  5342.70 & 4.02 & 0.574   \\   
	6449.82 & 2.52 & $-$0.630 & & 	  5483.36 & 1.71 & $-$1.220 \\  
	6455.60 & 2.52 & $-$1.370 & & 	   5647.23 & 2.28 &$-$1.580\\
	\multicolumn{3}{l}{\textbf{Sc} II ; $\log \epsilon_{\circ} = 3.10$ $A =$ 21}& & 6093.15 & 1.74 & $-$2.340 \\
	5239.82 & 1.45 & $-$0.760 & &    6455.00 & 3.63 & $-$0.280 \\
	5318.36 & 1.36 & $-$1.700 & & 	 \multicolumn{3}{l}{\textbf{Ni} I ; $\log \epsilon_{\circ} = 6.25$ $A = 28$} \\
	5526.82 & 1.77 & 0.150    & &    5578.72 & 1.68 & $-$2.650 \\
	6245.62 & 1.51 & $-$1.040 & &    5587.86 & 1.93 & $-$2.380 \\
	6300.69 & 1.51 & $-$1.960 & &    5682.20 & 4.10 & $-$0.390 \\
	6320.84 & 1.50 & $-$1.840 & &    5694.99 & 4.09 & $-$0.600 \\ 
	6604.60 & 1.36 & $-$1.160 & &    5805.22 & 4.17 & $-$0.580 \\ 
	\multicolumn{3}{l}{\textbf{Ti} I ; $\log \epsilon_{\circ} = 4.99$ $A =$ 22}& & 5847.00 & 1.68 & $-$3.410 \\
	5471.20 & 1.44 & $-$1.550 & &   6086.28 & 4.26 & $-$0.440 \\
	5474.23 & 1.46 & $-$1.360 & &   6111.07 & 4.09 & $-$0.800 \\
	5490.15 & 1.46 & $-$0.980 & &   6128.98 & 1.68 & $-$3.370 \\
	5866.46 & 1.07 & $-$0.840 & &   6130.14 & 4.26 & $-$0.950 \\
	6091.18 & 2.27 & $-$0.460 & &   \multicolumn{3}{l}{\textbf{Na} I ; $\log \epsilon_{\circ} = 6.33$ $A = 11$} \\
	6126.22 & 1.07 & $-$0.410 & &     5688.22 &  2.104 & $-$ 0.625 \\
	6258.11 & 1.44 & $-$0.440 & & 	  6154.23 &  2.102 & $-$ 1.607 \\					         
	6261.11 & 1.43 & $-$0.490 & &     6160.75 & 2.104  & $-$ 1.316  \\					        
	6303.76 & 1.44 & $-$1.600 & &    \multicolumn{3}{l}{\textbf{Mg} I ; $\log \epsilon_{\circ} = 7.58$ $ A= 12$} \\
	6312.24 & 1.46 & $-$1.580 & & 	  5711.09  &  4.346 & $-$ 1.706\\
	\multicolumn{3}{l}{\textbf{V} I ; $\log \epsilon_{\circ} = 4.00$  $A =$ 23}& &  6318.72 & 5.108  & $-$ 1.996 \\
	5668.37 & 1.08 & $-$1.00  & &    6319.24 & 5.108   & $-$ 2.179  \\
	5670.85 & 1.08 & $-$0.460 & &    8712.69 & 5.932   & $-$ 1.204  \\
	5727.05 & 1.08 & $-$0.000 & &    8736.02 & 5.946   & $-$ 0.224   \\
	5727.66 & 1.05 & $-$0.890 & &   \multicolumn{3}{l}{\textbf{Al} I ; $\log \epsilon_{\circ} = 6.47$ $A =13$} \\ 
	5737.07 & 1.06 & $-$0.770 & &    6696.03 & 3.143 & $-$ 1.570  \\					      
	6090.21 & 1.08 & $-$0.150 & &    6698.67 & 3.143 & $-$ 1.879  \\					      
	6216.35 & 0.28 & $-$0.900 & &    7835.31 & 4.022 & $-$ 0.728  \\					      
	6285.16 & 0.28 & $-$1.650 & &    7836.13 & 4.022 & $-$ 0.559  \\					     			 
	6452.31 & 1.19 & $-$0.820 & & 	 8772.87 & 4.022 & $-$ 0.425  \\
	 \multicolumn{3}{l} {}	  & &	 8773.91 & 4.022 & $-$ 0.212  \\

	\hline
   	\end{tabular}

\end{table}

\begin{table*}[!ht] 
\centering
%\footnotesize
\caption{Spectrographs and data characteristics.} 
\begin{tabular}{|l|c|c|c| } 
\hline Spectrograph/Telescope            & Observatory &Resolution & Range\\ 
&  &($\lambda$/$\Delta\lambda$) & (\AA)\\ 
\hline 
\hline 
CORALIE/1.2-m Ewler Swiss & La Silla(Chile) & 50000 & 3800-6800 \\ 
FEROS/1.52-m ESO        & La Silla (Chile)  & 48000  & 3600-9200 \\ 
UVES/VLT 8-mKuyen UT2  & Paranal (Chile)   & 110000 & 4800-6800 \\ 
SARG/3.5-m TNG          & ORM (la Palma)    & 57000  &5100-10100\\ 
UES/4-m WHT             & ORM (la Palma)    & 55000  & 4600-7800\\    
\hline 
\end{tabular}
\label{strum} 
\end{table*}

\section{Data}
All the objects from the comparison sample belong to the CORALIE extrasolar-planets-finding 
programme \footnote{http://unige.ch/~udry/planet/planet.html.}.
The high resolution spectra are the same as those used by Santos et
al.~(2004a) to derive precise and uniform stellar parameters for 98 planet-host stars and 41 comparison sample ``single'' dwarfs in a 
volume-limited sample in the solar neigborhood ($<$ 20 pc).  It should be pointed out that the star HD~219542~B was excluded from the 
planet-host list presented in Santos
et al.~(2004a) since the presence of a planet around this star has been rejected (Desidera et al.~2004).
All our spectra were collected during several observing campaigns with the CORALIE spectrograph on the 1.2 m Euler
Swiss Telescope, the FEROS spectrograph on the 2.2 m ESO/MPI Telescope (both at La Silla, Chile), the UVES spectrograph on the VLT/UT2
Kueyen Telescope (Paranal Observatory, ESO, Chile), the SARG spectrograph on the 3.5 m TNG
and the UES spectrograph on the 4.2 m WHT (both at the Roque de los Muchachos Observatory, La Palma, Spain). The spectrograph and data
characterisctics are listed in Table~\ref{strum}.\\
In order to extend the number of stars in a comparison sample, we added 48 FEROS spectra from Santos et
al. (2005) and five new SARG spectra. We refer the reader to this paper for a description of the data.
The {S/N} ratio of the spectra varies between 150 and 350. New high {S/N} spectra obtained in 2004 with VLT/UVES have been
used as well (see Ecuvillon et al.~2004a,2004b, 2005)

\begin{figure*}[tb]
\begin{tabular}{lr}
%\centering
\psfig{width=0.5\hsize, file=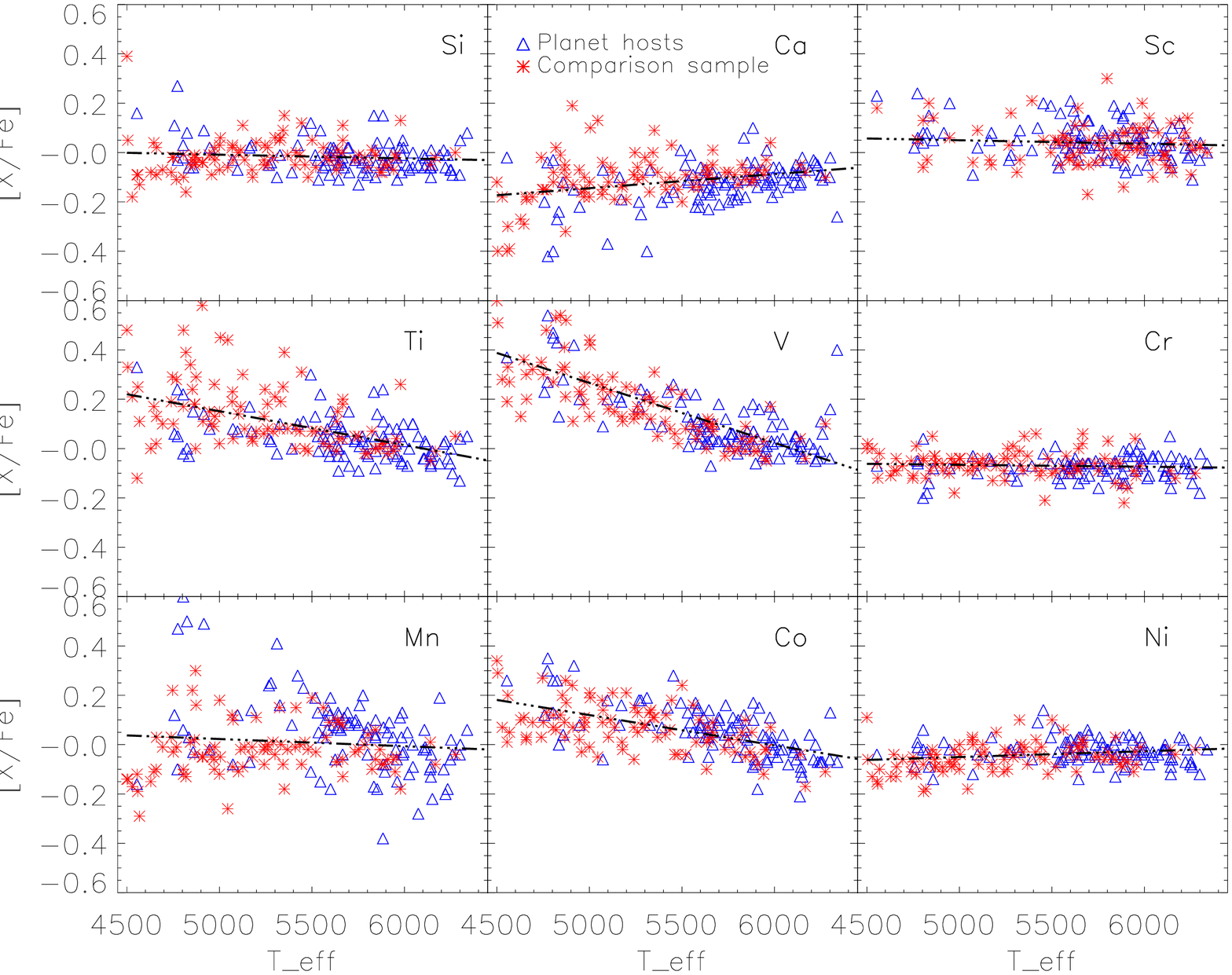}&\psfig{width=0.5\hsize, file=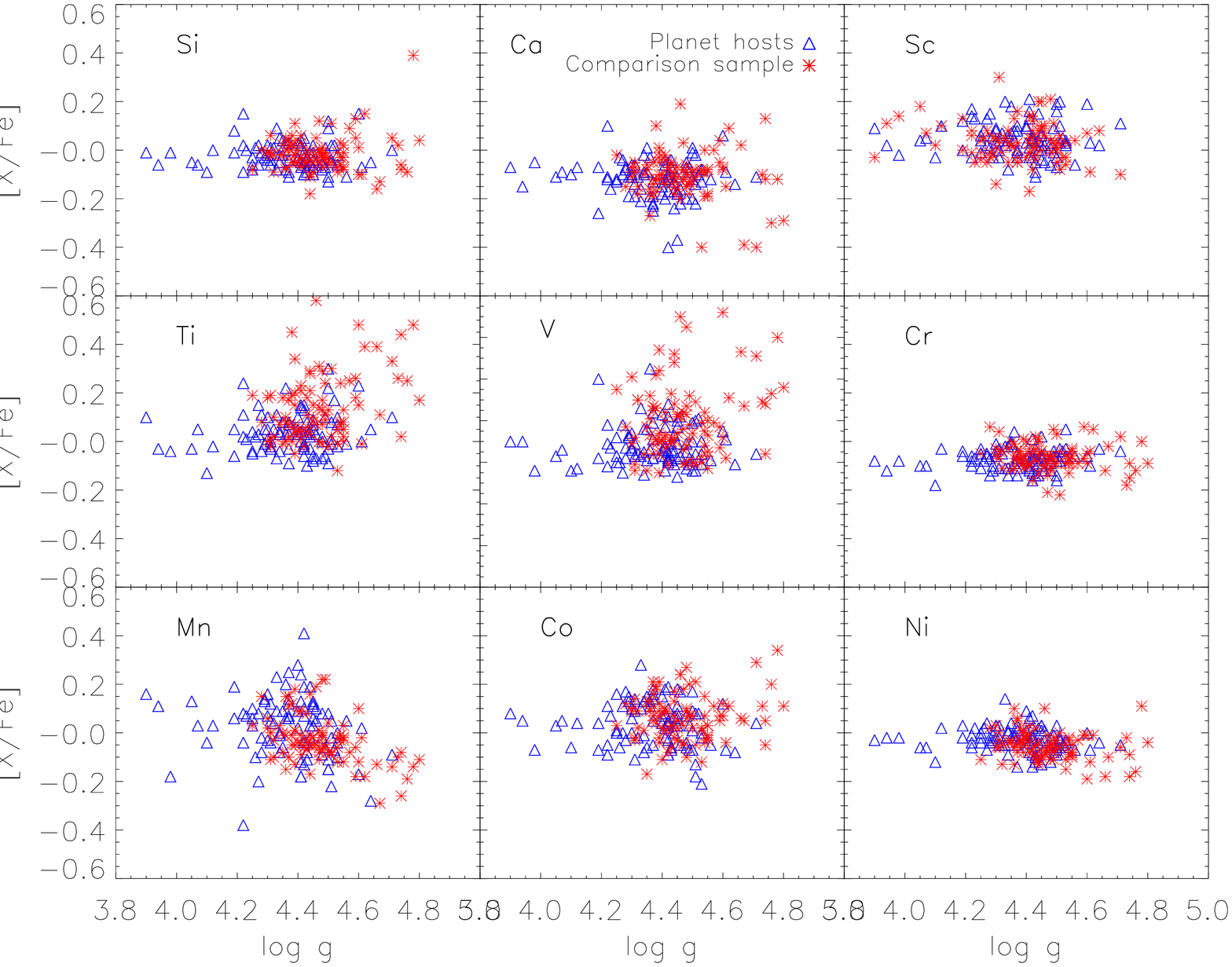}\\
\end{tabular}
\caption{[$X$/Fe] {vs.}\ $T_{\rm eff}$ and [$X$/Fe] {vs.}\ $\log g$. Triangles and asterisks are planet host and 
comparison
sample stars, respectively. Dotted lines represent the solar value. The slopes represent linear least-squares fits to both
planet-hosts and comparison stars.}
\label{slope}
\end{figure*}

\begin{figure}[!h]
%\centering
\psfig{width=\hsize, file=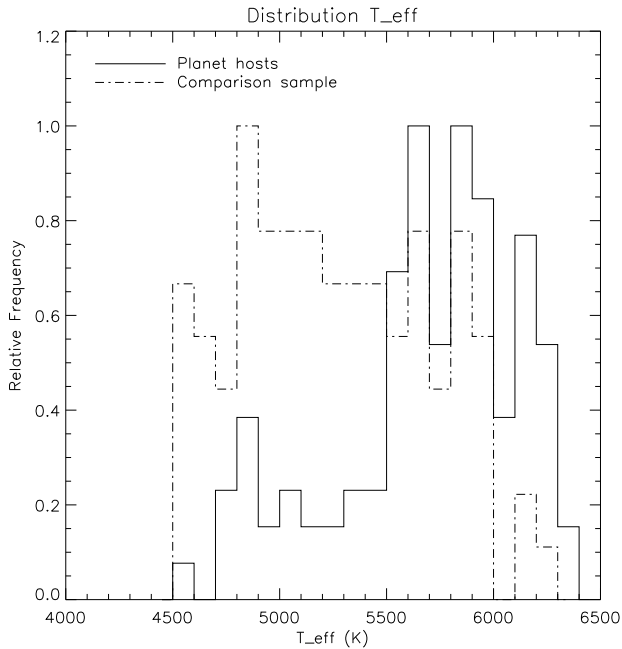}
\caption{Distributions of planet-host stars and non-planet-host stars as functions of $T_{\rm eff}$. The planet-host stars and the
 comparison sample are represented with solid and dotted lines, respectively.}
\label{istemp}
\end{figure}

\begin{figure}[!tb]
%\centering
\psfig{width=\hsize, file=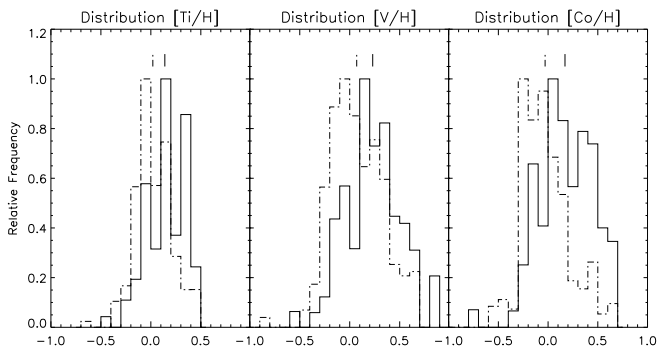}
\caption{[$X$/H] distributions for Ti, V and Co for the two subsamples of planet host (solid line) and
comparison sample (dotted lines) stars with the \textit{same} $T_{\rm eff}$ distributions.}
\label{istotest}
\end{figure}

\section{Spectral analysis}
Chemical abundances of the nine refractory elements  studied here, as well as Na, Mg and Al, were derived performing a standard local 
thermodynamic equilibrium (LTE)
analysis, strictly differential with respect to the Sun. Solar abundances for each element were taken from Anders \& Grevesse (1989) 
using a solar model with $T_{\rm eff}$ = 5777 K,
$\log g$ = 4.44 dex, $\xi_{t}$  = 1.0 km s$^{-1}$.
We used a revised version (2002) of the MOOG code (Sneden 1973)\footnote{The MOOG2002 source code can be downloaded at
http://verdi.as.utexas.edu/moog.html.} (with the \textbf{abfind} driver) and a grid of Kurucz (1993) ATLAS9 atmospheres.
The atmospheric parameters, effective temperature ($T_{\rm eff}$), surface gravity ($\log g$), microturbolence ($\xi_{t}$) and
metallicity ([$Fe$/$H$]), and their corresponding uncertainties, were taken from Santos et al.~(2004a, 2005). 
It should be stressed that these stellar parameters were derived in a uniform way, using high resolution spectra, the same
line list and model atmospheres for all the stars.
 The spectral lines of the refractory elements analysed here were extracted from the study by Bodaghee et al.~(2003), while the 
 line lists
 of Na, Mg and Al were taken from Beir{\~a}o et al.~(2005). These lists were successively modified in order to minimize abundance 
 errors. Since the  V~I line at 6531.42 \AA , the 
 Mn~I line at 5388.50 \AA \,and the Co~I lines at 5312.86 \AA \ and 6632.44 \AA\ were difficult to measure in most of the
 spectra (too weak or blended), we eliminated them from the list. We also excluded the Mg line at 8923.57 \AA\ 
 because it
 was not measurable in the new Feros spectra analysed for Na, Mg and Al.
  We added more lines of V, Mn and Co from  Gurtovenko \& Kostik (1989). Before including these lines in our list, we first 
  verified that each
 line was not too strong and checked for
 possible blending, using the Kurucz Solar Atlas (Kurucz et al.~1984).
For these new lines we derived semi-empirical atomic oscillator strengths using their EWs measured in the solar atmosphere with 
($T_{\rm eff}$,
$\log g$, $\xi_{t}$)= 5770 K, 4.44 dex and 1.0 km $s^{-1}$), and performed an
inverted solar analysis. The number of lines increased from 5 to 9 for V, from 3 to 8 for Mn, and from 5 to 7 for Co.

Finally, we considered about 80 spectral lines for the analysis of refractory elements in each spectrum and about 14 lines for the 
study of Na, Mg and Al
abundances (see Table~\ref{linee}) in 53 new FEROS and SARG spectra. We have measured about 16\,600 equivalent widths
 in our spectra. For each spectral line EWs were measured  by a Gaussian fit using the \textit{splot} task within the 
``echelle'' IRAF
package.\footnote{IRAF is distributed by National Optical Astronomy Observatories, operated by the Association of Universities for
Research in Astronomy, Inc., under contract with the National Science Fundation, USA.}
Furthermore we employed a new program (IRAF scripts) written by N.~C.~Santos for automatic measure of the EWs of Na, Mg and Al
lines.
This program was previously used to measure the EWs of a list of Fe lines (Santos et al.\ 2005). Every measurement was taken carefully 
achieving the best agreement between lines profile and Gaussian fits.
About 25\% of the equivalent widths used to calculate abundances come from Bodaghee et al.~(2003), whose stellar parameters
have been updated by Santos et al.\ (2004a, 2005). Detailed observational data (e.g.\ line-by-line EWs for each element) are available as
an electronic table (Table 11) at the Centre des Donn\'ees Strasbourg (CDS).
\label{anal}

\subsection{Uncertainties}
\label{erro}
We have tested the dependence of our results on atmospheric parameters (Fig.~\ref{slope}) and
refer the reader to the Figs~1--2 by Beir{\~a}o et al.\ (2005) for the plots of [$X$/Fe] ($X$ = Na, Mg, Al) {vs.}\ $T_{\rm eff}$ and {vs.}\ $\log g$. In 
Table~\ref{tabNLTE} are listed the slopes of the [$X$/Fe] ratios for refractory elements as a function of $T_{\rm eff}$ for all stars in the distributions. We observe no characteristic trends for the majority of the elements studied.
In the case of Ti, V and Co we note decreasing trends for the [$X$/Fe] ratio corresponding to  $T_{\rm eff}$  values greater than 5500
K. This trend could be caused by NLTE effects which, however, contribute mostly in the case of the more metal-poor
stars. We do not take NLTE effects into account because they are usually small in metal-rich stars for complex atoms such as 
iron (see Edvardsson et al.~1993; Th{\'e}venin \& Idiart 1999).
On the other hand, although NLTE effects have not been studied for a number of elements (e.g.\ V, Co, etc.), they are nonetheless not thought to alter the main
conclusion of the present paper. 
Some unknown blended lines may also be responsible for these trends. In fact, the overestimation of
EWs caused by the increasing blending effects becomes more severe at lower $T_{\rm
eff}$.

The distribution of planet-hosts and non-planet-host stars as function of $T_{\rm eff}$ 
shows that the latter are on average cooler (see
Fig.~\ref{istemp}). 
To verify this, we selected randomly two subgroups of stars (planet and no-planet hosts) with the same $T_{\rm eff}$
distributions and obtained the [$X$/H] distribution for Ti, V and Co. We repeated this process for 200 randomly selected subgroups and 
calculated the average distribution for each element (Ti, V, Co). In Fig.~\ref{istotest} we observed the same behaviour as that found
for all the stars analysed. In both cases the differences between average [$X$/H] for hosts and no-hosts are of the same order (see
Table~\ref{tabmed}.

%Different types of errors may affect abundance measurements. Errors in the EWs may be caused by unknown blends or poor location 
%of the continuum. Some authors (e.g.\ Santos, Israelian \& Mayor 2001) 
%have evaluated the errors caused by EWs measured by two independent observers and have compared values obtained for selected Fe I and Fe
%II lines. They found no systematic differences: the two samples presented a mean deviation of only 3 m\AA, corresponding to an
%uncertainty of 0.05 dex in the resulting abundances calculated by  two different observers. This comparison suggests that we did not introduce
%significant errors into the calculation of abundances using  25\%  EW values by Bodaghee et al.\ (2003).

\begin{table}[tb]%TA1
\centering
\par
	\caption{ Slopes of [$X$/Fe] ratios as functions of effective temperatures per 1000 K (in dex) for both samples (planet hosts and
	non-hosts).}
	\label{tabNLTE}
   	\begin{tabular}{lr|lr}	
	\hline
	Species	&	Slopes $\pm$ rms  &  Species	&	Slopes $\pm$ rms       \\
	\hline\hline
	\noalign{\smallskip}
	\noalign{\smallskip}
	
	Si	&	$-0.015 \pm 0.010$ &   Cr      &       $-0.007 \pm 0.007$   \\
	Ca	&	$0.057 \pm 0.013$  &   Mn      &       $-0.029 \pm 0.020$  \\
	Sc	&	$-0.014 \pm 0.011$ &   Co      &       $-0.122 \pm 0.012$  \\
	Ti     	&	$-0.137\pm 0.015$  &   Ni      &       $ 0.023 \pm 0.008$  \\
	V	&	$-0.242 \pm 0.015$   &  \\
	   
	  \hline
   	\end{tabular}
\end{table}

In contrast, systematic errors are difficult to locate but are largely reduced by good data
quality and good instrumental resolution. Since we have analysed more than one line for each element (usually 7--8 lines) in a 
given star, the total dispersions 
around the average abundance are more significant compared to the continuum observed uncertainties, which are usually
around 0.05 dex.

\begin{table*}[bt]
\caption{\small{Sensitivity of the abundances of  Si, Ca, Sc, Ti, V, Cr, Mn, Co and Ni to changes of 100 K in $T_{\rm eff}$, 0.30
dex in $\log g$ and [Fe/H], 0.50 km $s^{-1}$ in $\xi_{t}$}.}
\centering
%\footnotesize
\begin{tabular}{|l|c|c|c|c|c|c|c|c|c| }
\hline
\noalign{\smallskip}
Star     &    Si & Ca & Sc & Ti & V& Cr & Mn & Co & Ni  \\
($T_{\rm eff}$;[$Fe$/$H$];$\log g $;$\xi_{t}$) & & & & & & & & & \\
\noalign{\smallskip}
\hline
\hline
\multicolumn{10}{ @ {}c @ {}}{Temperature variation $\Delta T_{\rm eff}$ $\pm$ 100 K}  \\
\hline
\noalign{\smallskip}
HD 50281A & $\mp$0.06 & $\pm$0.12 & $\mp$0.01 & $\pm$ 0.14 &  $\pm$ 0.15    & $\pm$ 0.06   & $\pm$ 0.05 & $\pm$ 0.01&$\mp$0.01\\
(4685;$-$0.04;4.32;0.64) & & & & & & & & & \\
\hline
HD 43162 & $\pm$ 0.01 & $\pm$ 0.08 & $\mp$0.01 & $\pm$ 0.10 &  $\pm$ 0.11 &  $\pm$ 0.06   & $\pm$0.06 & $\pm$0.06 & $\pm$ 0.05  \\
(5633;$-$0.01;4.48;1.24 )& & & & & & & & & \\
\hline
HD 10647 &  $\pm$ 0.03 & $\pm$ 0.07 &  $\mp$ 0.01 & $\pm$ 0.09 & $\pm$ 0.10    & $\pm$ 0.05  &$\pm$ 0.06 & $\pm$ 0.07 &  $\pm$ 0.06\\
(6143;$-$0.03;4.48;1.40) & & & & & & & & & \\
\hline
\multicolumn{10}{ @ {}c @ {}}{Surface gravity variation $\Delta$ $\log g$ $\pm$ 0.30 dex} \\
\hline
HD 10697 & $\pm$ 0.00 &  $\mp$0.06 & $\pm$ 0.11 & $\mp$0.01& $\mp$ 0.01     & $\mp$0.01  &$\mp$ 0.02 & $\pm$0.01 & $\pm$0.00 \\
(5641;0.14;4.05;1.13)& & & & & & & & & \\
\hline
HD168443 & $\pm$0.01 & $\mp$0.07 & $\pm$0.12 & $\mp$ 0.01& $\mp$ 0.00     & $\mp$ 0.02  & $\mp$0.02 & $\pm$0.01 & $\pm$0.00 \\
(5617;0.06;4.22;1.21)& & & & & & & & & \\
\hline
HD 28185 & $\pm$ 0.00 & $\mp$0.09 & $\pm$ 0.11 & $\mp$ 0.02& $\mp$ 0.02     & $\mp$ 0.03  & $\mp$0.04 &$\pm$0.02 & $\pm$ 0.01 \\
(5656;0.22;4.45;1.01)& & & & & & & & & \\
\hline
\multicolumn{10}{ @ {}c @ {}}{Metallicity variation $\Delta$ [$Fe$/$H$] $\pm$0.30 dex} \\
\hline
HD  6434  & $\pm$ 0.01 & $\pm$ 0.00 & $\pm$0.06 &  $\mp$0.01& $\mp$0.01     & $\pm$0.00  & $\pm$0.00 & $\pm$0.00 &  $\pm$0.00  \\
(5835;$-$0.52;4.60;1.53)& & & & & & & & & \\
\hline
HD147513  & $\pm$ 0.00 & $\pm$ 0.02 & $\pm$ 0.08 &$\mp$0.01& $\mp$0.01     & $\pm$ 0.00  & $\pm$ 0.00 & $\pm$0.01 & $\pm$ 0.01 \\
(5894;0.08;4.43;1.26)& & & & & & & & & \\
\hline
HD  4203  & $\pm$ 0.00 & $\pm$ 0.03 & $\pm$ 0.10 & $\mp$0.00& $\mp$0.01     & $\pm$ 0.01  & $\pm$ 0.02 &$\pm$ 0.03 & $\pm$ 0.04 \\
(5636;0.40;4.23;1.12& & & & & & & & & \\
\hline
\multicolumn{10}{c}{Microturbulence variation $\Delta$ $\xi_t$ $\pm$0. 50 km $s^{-1}$} \\
\hline
HD 69830 & $\mp$0.02 & $\mp$0.07 &$\pm$ 0.01 & $\pm$ 0.00 & $\mp$0.07    & $\mp$0.04   & $\mp$0.08 & $\mp$0.06&$\mp$0.06\\
(5410;$-$0.03;4.38;0.89) & & & & & & & & & \\
\hline
HD 43162 &  $\mp$ 0.02 & $\mp$ 0.08 & $\mp$ 0.06 & $\mp$ 0.05 &  $\mp$0.04    & $\mp$0.04   & $\mp$0.08 & $\mp$0.05 & $\mp$0.05 \\
(5633;$-$0.01;4.48;1.24 )& & & & & & & & & \\
\hline
HD 84117 &  $\mp$0.02 & $\mp$ 0.08 &  $\mp$ 0.06 & $\mp$ 0.05 &  $\mp$ 0.04    & $\mp$ 0.05   &$\mp$0.05 & $\mp$ 0.02 &  $\mp$0.04\\
(6167;$-$0.03;4.35;1.42 )& & & & & & & & & \\
\hline
\end{tabular}
\label{sens}

\end{table*}

When analysing many lines to calculate abundances, uncertainties in the atmospheric parameters should be the primary source of abundance errors.
These are of the order of 50 K in $T_{Â\rm eff}$, 0.12 dex in $\log g$, 0.08 km $s^{-1}$ in $\xi_{t}$ and 0.05 dex in [Fe/H] (see
Santos et al.~2004b, 2005).
The abundance sensitivity to changes in atmospheric parameters were estimated as follows. First we selected a set of 12 stars 
from our list: for each  atmospheric parameter we chose three stars with similar values for all the parameters except for the one considered, which must vary within the sample. We then generated new atmospheric models, changing only the ``varying'' parameter and calculating the
corresponding abundances. The difference between these values and those obtained without varying the parameter, give us the abundance
sensitivity to changes in the parameters. Tables~\ref{sens} and~\ref{sens2} show these results after varying the effective temperature by 
$\Delta T_{\rm eff}= \pm 100 $ ,
surface gravity by $\Delta \log g= \pm 0.30 \, dex$, metallicity by $\Delta [$Fe$/$H$]= \pm 0.30 \,dex$  and microturbulence by 
$\Delta \xi_t = \pm 0.50 \, km\,s^{-1}$. We note that ions such as Sc II are generally more sensitive to changes in surface gravity,
while neutral atoms are influenced mostly by uncertainties in the effective temperature. At a glance, we observe from  
Table~\ref{sens} that the V and Ti
abundances can vary with temperature changes of $\Delta[$X$/$H$]\sim \pm 0.12 \, dex$, while we associate 
$\Delta [Sc/H]\sim \pm 0.11 \, dex$
with changes in surface gravity. With respect to Table~\ref{sens2} we note that Na abundances are more sensitive to variations in
atmospheric parameters (e.g.\ $T_{Â\rm eff}$ and $\log g$) than Mg and Al.
Finally, we evaluated the errors in the abundances of all the elements, adding quadratically the standard deviation of the mean abundance obtained from all the measured lines and the uncertainties dued to the abundance sensitivities to changes in the atmospheric parameters. For each star these "total" errors are of the order of 0.10 dex.

\begin{table}[hbt]
\caption{\small{The same as Table~\ref{sens} but for  Na, Mg and Al.}}
\centering
%\footnotesize
\begin{tabular}{|l|c|c|c|}
\hline
\noalign{\smallskip}
Star     &    Na & Mg & Al  \\
($T_{\rm eff}$;[$Fe$/$H$];$\log g $;$\xi_{t}$) & & &  \\
\noalign{\smallskip}
\hline
\hline
\multicolumn{4}{ @ {}c @ {}}{Temperature variation $\Delta T_{\rm eff}$ $\pm$ 100 K}  \\
\hline
\noalign{\smallskip}
HD 50281A & $\pm$0.09 & $\pm$0.00 & $\pm$0.07 \\
(4685;$-$0.04;4.32;0.64) & & & \\
\hline
HD 43162 &  $\pm$ 0.05 & $\pm$0.05 & $\pm$0.05  \\
(5633;$-$0.01;4.48;1.24 )& & & \\
\hline
HD 10647 &  $\pm$0.05 & $\pm$0.05 &  $\pm$0.05 \\
(6143;$-$0.03;4.48;1.40) & & &  \\
\hline
\multicolumn{4}{ @ {}c @ {}}{Surface gravity variation $\Delta$ $\log g$ $\pm$ 0.30 dex} \\
\hline
HD 10697 & $\mp$0.12 &  $\mp$0.05 &$\mp$ 0.02 \\
(5641;0.14;4.05;1.13)& & & \\
\hline
HD168443 & $\mp$0.11 &  $\mp$0.05 & $\mp$0.03 \\
(5617;0.06;4.22;1.21)& & &  \\
\hline
HD 28185 & $\mp$0.05 &  $\mp$0.03 & $\mp$0.03  \\
(5656;0.22;4.45;1.01)& & &  \\
\hline
\multicolumn{4}{ @ {}c @ {}}{Metallicity variation $\Delta$ [$Fe$/$H$] $\pm$0.30 dex} \\
\hline
HD  6434  & $\mp$0.01 & $\mp$0.04 & $\mp$0.01   \\
(5835;$-$0.52;4.60;1.53)& & &  \\
\hline
HD147513  & $\pm$0.02 & $\mp$ 0.04 &  $\pm$0.00  \\
(5894;0.08;4.43;1.26)& & &  \\
\hline
HD  4203  & $\pm$0.02 & $\pm$0.06 &$\pm$ 0.00 \\
(5636;0.40;4.23;1.12 & & & \\
\hline
\multicolumn{4}{c}{Microturbolence variation $\Delta$ $\xi_t$ $\pm$0.50 km $s^{-1}$} \\
\hline
HD 69830 & $\mp$0.02 & $\mp$0.04 & $\mp$0.02 \\
(5410;$-$0.03;4.38;0.89) & & & \\
\hline
HD 43162 &  $\mp$0.02 & $\mp$0.04 & $\mp$0.02 \\
(5633;$-$0.01;4.48;1.24 )& & &  \\
\hline
HD 84117 & $\pm$ 0.02 & $\pm$0.04 &  $\mp$ 0.03 \\
(6167;$-$0.03;4.35;1.42 )& & & \\
\hline
\end{tabular}
\label{sens2}

\end{table}

\section{Results}

\begin{figure*}[!htb]
\begin{center}
\psfig{width=0.8\hsize,file=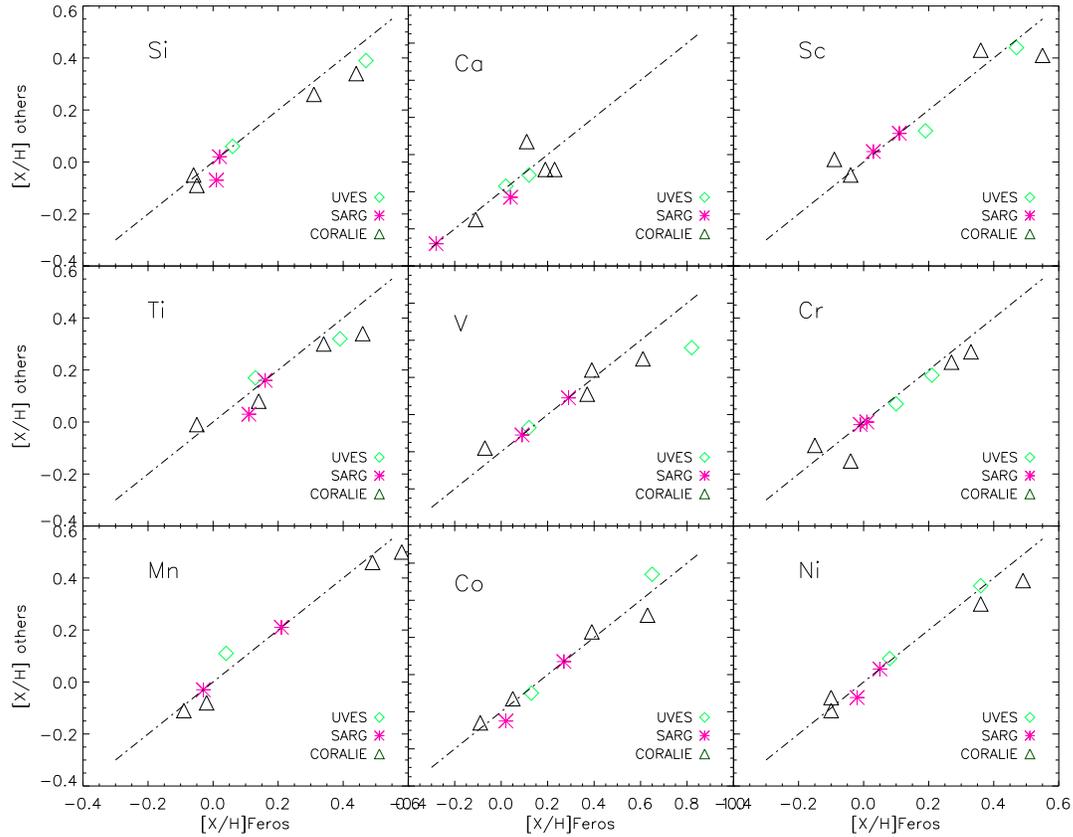}
\caption{\small{Comparison of the abundances of nine refractory elements calculated with EW values measured from spectra taken with different instruments.
Along the $x$-asis are results from FEROS spectra, while along the $y$-axis we represent abundance values for the same objects, 
calculated using
other spectra. UVES, CORALIE and SARG results are respectively represented with rhombi (HD~16141, HD~27442), triangles (HD~83443,
HD~92788, HD~106252, HD~114386) and asterisks (HD~114783, HD~147513).}}
\label{compa}
\end{center}
\end{figure*}

Bodaghee et al.~(2003) recently carried out a spectroscopic analysis of the same refractory elements as those presented in our
work while Beir{\~a}o et al.\ (2005) did the same for Na, Mg and Al. These authors did not find any significant differences
 between planet host and 
comparison-sample stars, a result in perfect agreement with
our findings. Because of the lack of 
comparison sample stars with [Fe/H] $>0.1$ dex in previous studies, the abundance distribution of stars with giant planets
looked like a high metallicity extension
of the curves traced by field dwarfs without planets. New spectra of metal-rich stars with no planets have been
gathered, and consequently  a complete comparison is also possible in the high metallicity domain. It is still conceivable that certain specific 
trends found for the metal-rich tail of  Galactic chemical evolution are due to the presence of planets (see plots for Co, V, Na,
Mg, Al in Figs ~\ref{fig1},~\ref{fignew}). However, we should stress that even the new comparison sample contains about 
20\% of stars at [Fe/H] $>0.1$.
All the results, together with their total errors (see Sec.~\ref{erro}), are listed in the Tables~7-10.
Furthermore, we have compared the abundances calculated with EW values measured with different instruments. As shown
in Fig.~\ref{compa}, we did not detect any significant discrepancies among the abundance values for the nine refractory elements.

\begin{figure*}[!htb]
\begin{tabular}{lcr}
\psfig{width=0.31\hsize, file=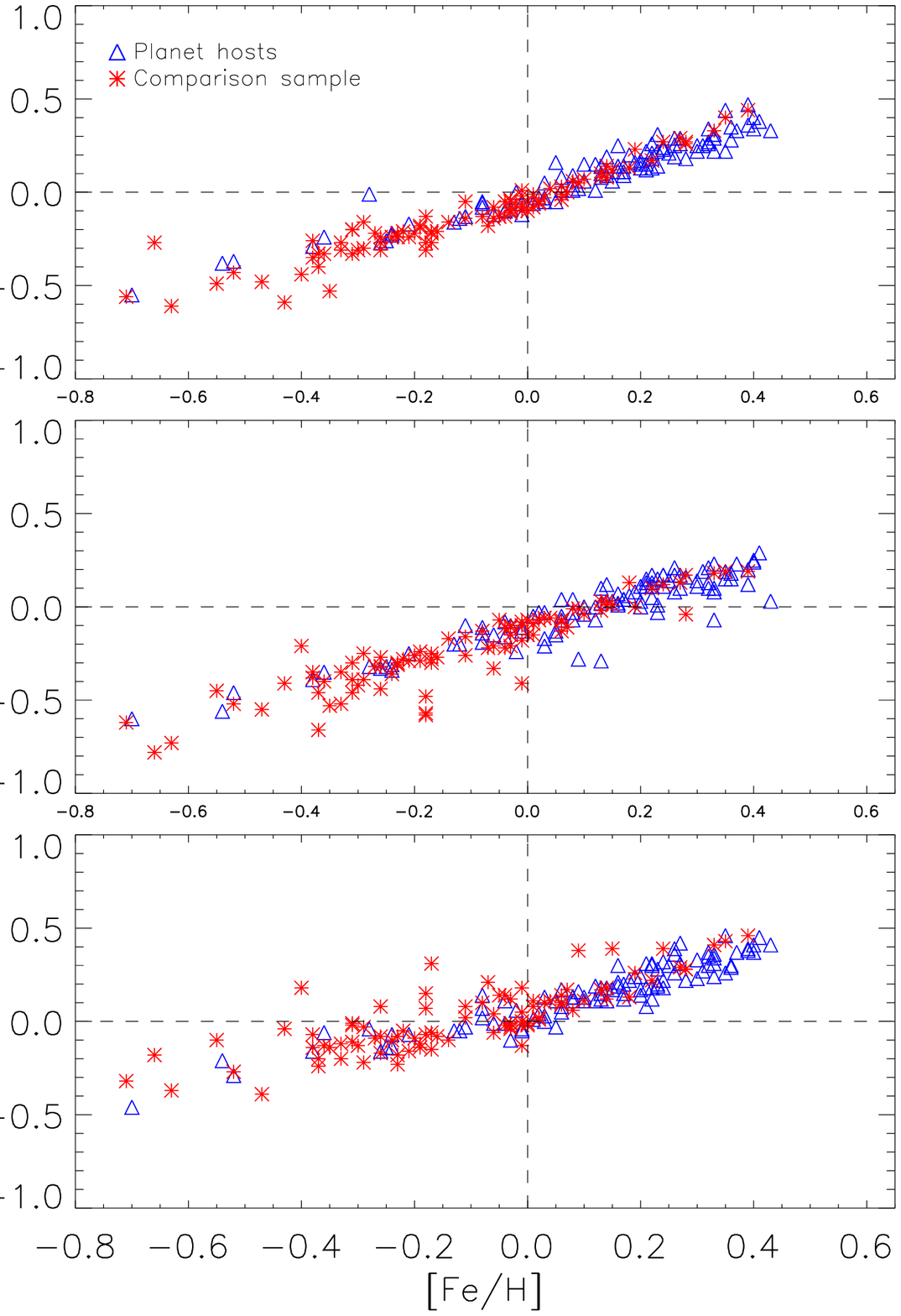}&\psfig{width=0.31\hsize, file=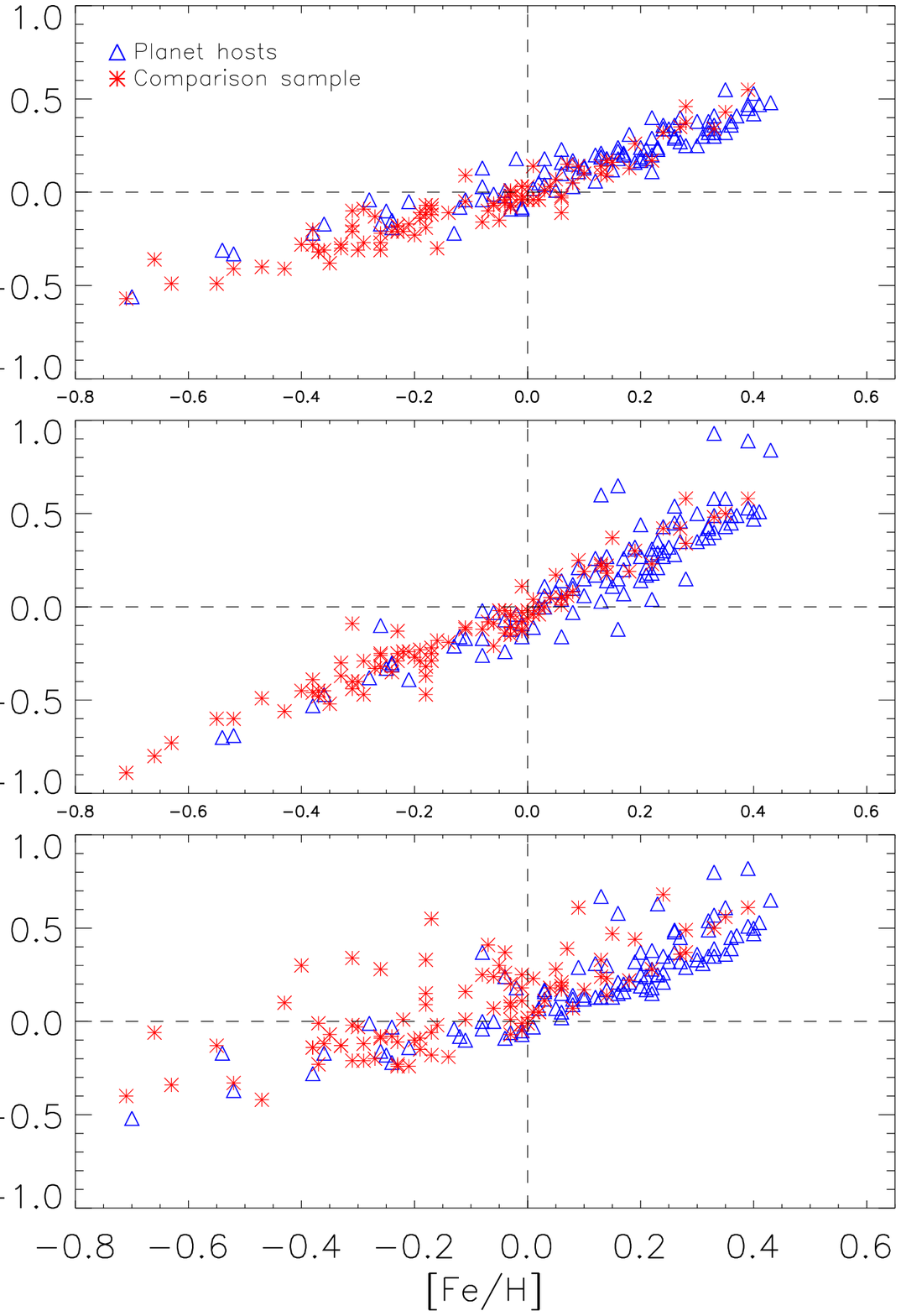} &\psfig{width=0.31\hsize,file=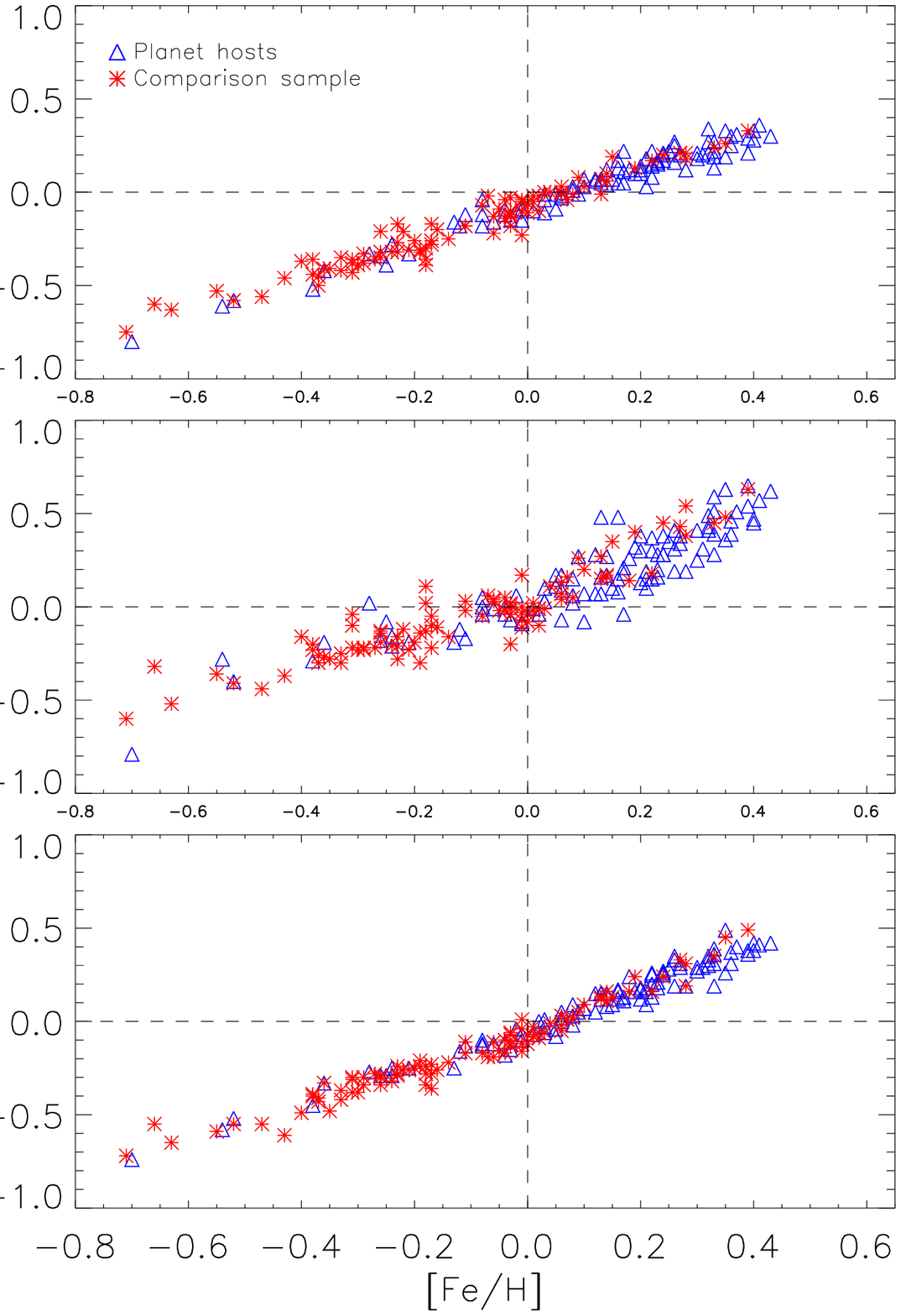}\\
\end{tabular}
\caption{[$X$/H] {vs.}\ [Fe/H] for Si, Ca, Ti (left), for Sc, Mn and V (centre), for Cr, Co and Ni (right). Triangles 
and asterisks represent stars with planetary-mass
companion and ``single'' field stars, respectively. The intersection of the dotted lines indicates solar value.}
\label{fig0}
\end{figure*}

%\newpage

\begin{figure*}[htb]
\centering
\psfig{width=0.9\hsize,file=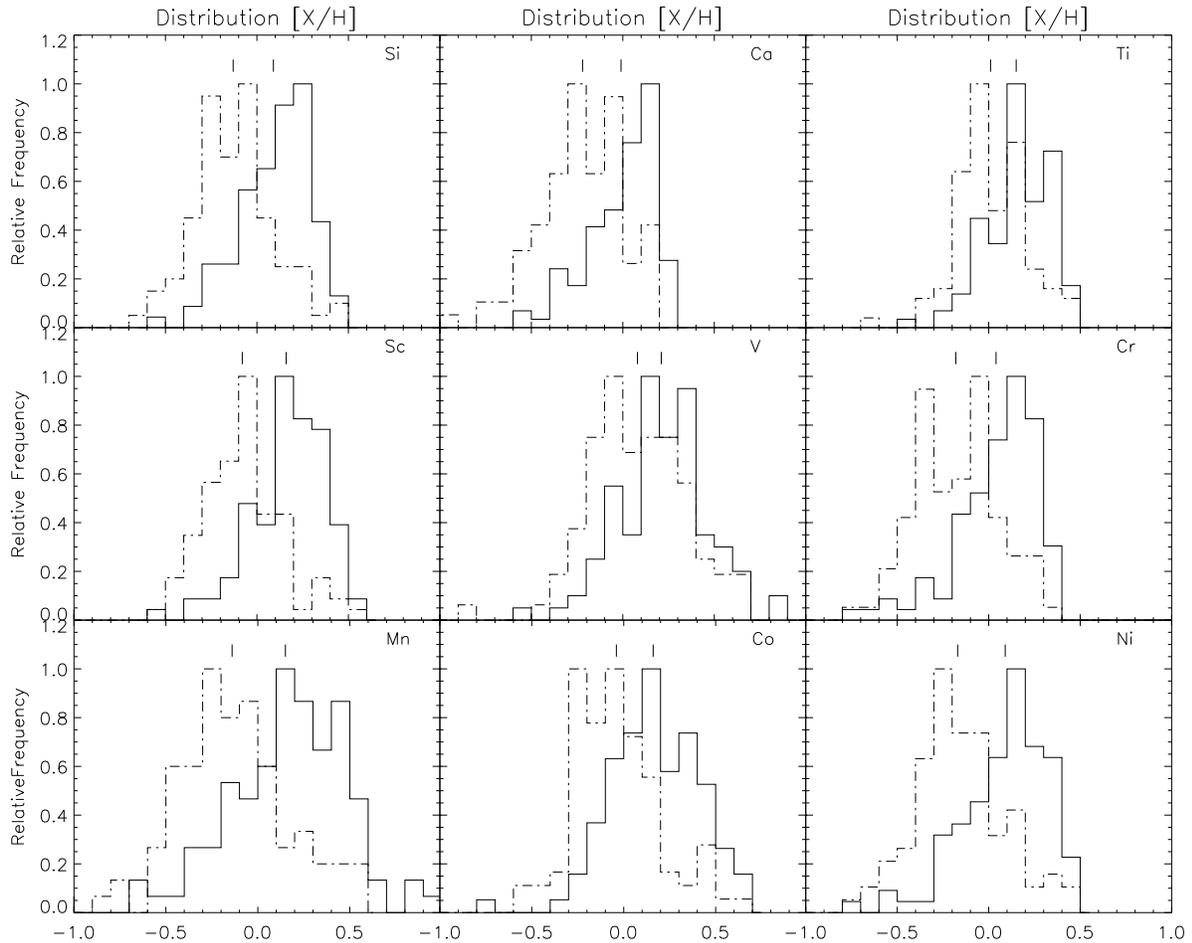}
\caption{[$X$/H] distribution for each element. The planet-host stars and the comparison sample are represented with solid and 
dotted lines, respectively.}
\label{isto}
\end{figure*}

\begin{table}[b]
%\centering
\caption{\small{Average abundances $<$[$X$/H]$>$ for stars with planets and for comparison sample stars.
For each element the rms around the mean and the abundances difference between the two samples (101 and 94 respectively) are also listed.}}
\begin{tabular}{lccccl}
\noalign{\smallskip}
\noalign{\smallskip}
 \hline
Species &       \multicolumn{2}{c}{Planet-hosts} & \multicolumn{2}{c}{No-hosts}   & Average   \\
        (X) &       $<$[$X$/$H$]$>$ & $\sigma$ & $<$[$X$/$H$]$>$ & $\sigma$ &  diff.   \\
        \noalign{\smallskip}
        \noalign{\smallskip}
        \hline
	\hline
        \noalign{\smallskip} 
        Si      &       0.09 & 0.20 & $-$0.13   & 0.24 &0.23\\
        Ca      &      $-$0.01 & 0.19 & $-$0.22 & 0.22&0.22\\
        Ti      &       0.15 & 0.17 & 0.01      & 0.19  &0.14\\
        Sc      &       0.16 & 0.20 & $-$0.08   & 0.24 &0.25\\
        V       &       0.21 & 0.24 & 0.08     & 0.26  &0.13\\
        Cr      &       0.04 & 0.22 & $-$0.18  & 0.25 &0.22\\
        Mn      &       0.15 & 0.33 & $-$0.14  & 0.33 &0.29\\
        Co      &       0.16 & 0.25 & $-$0.04  & 0.26 &0.20\\
        Ni      &       0.09 & 0.24 & $-$0.17  & 0.27 &0.26\\
        Na      &       0.12 & 0.24 & $-$0.17  & 0.26 &0.29\\
	Mg	&	0.19 & 0.19 & $-$0.05  & 0.21 &0.23\\
	Al	&	0.19 & 0.21 & $-$0.01  & 0.23 &0.18\\
	
	\hline
\end{tabular}
\label{tabmed}
\end{table}

\begin{figure*}[!ht]
\begin{tabular}{lr}
\psfig{width=0.45\hsize, file=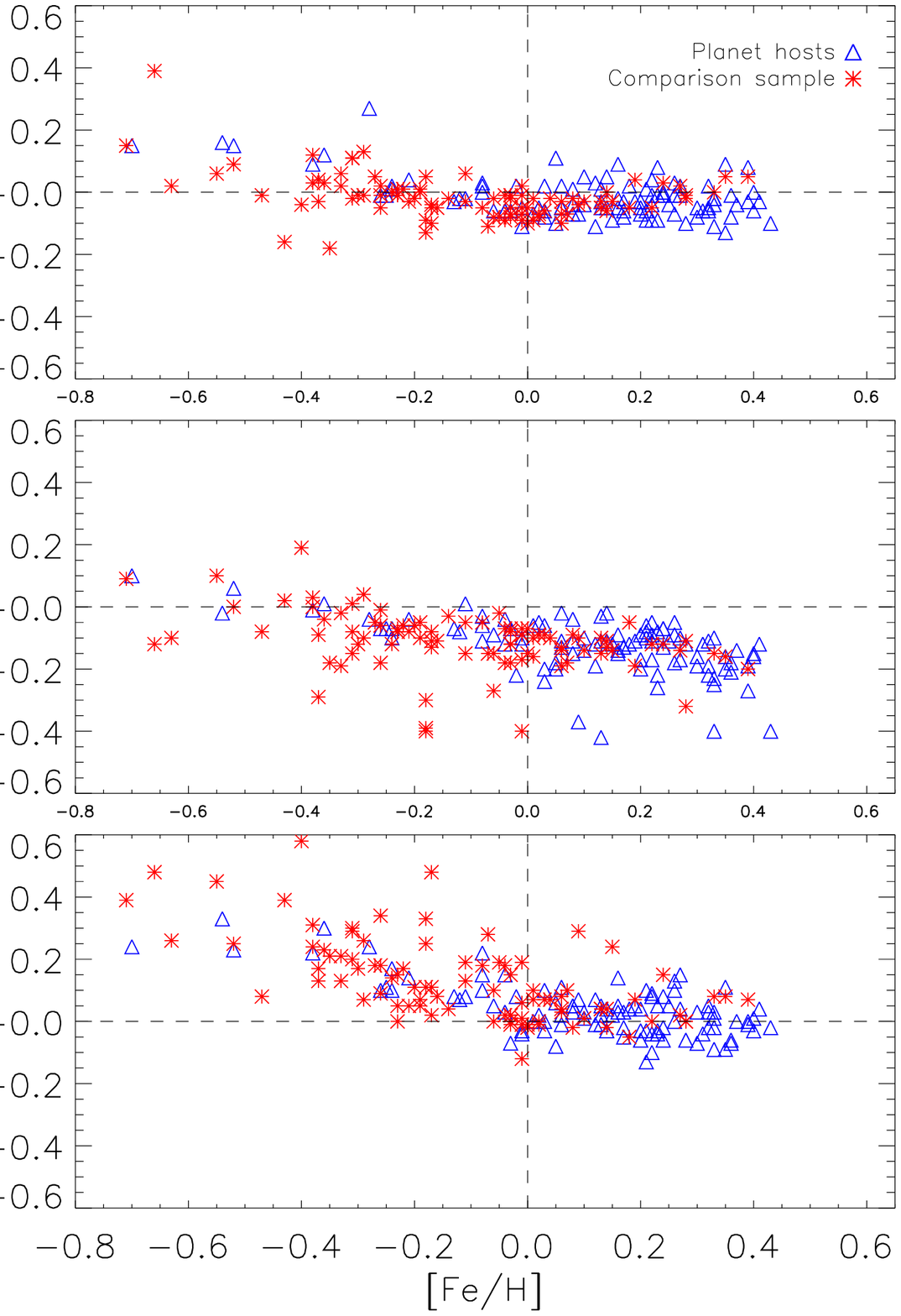}&\psfig{width=0.45\hsize, file=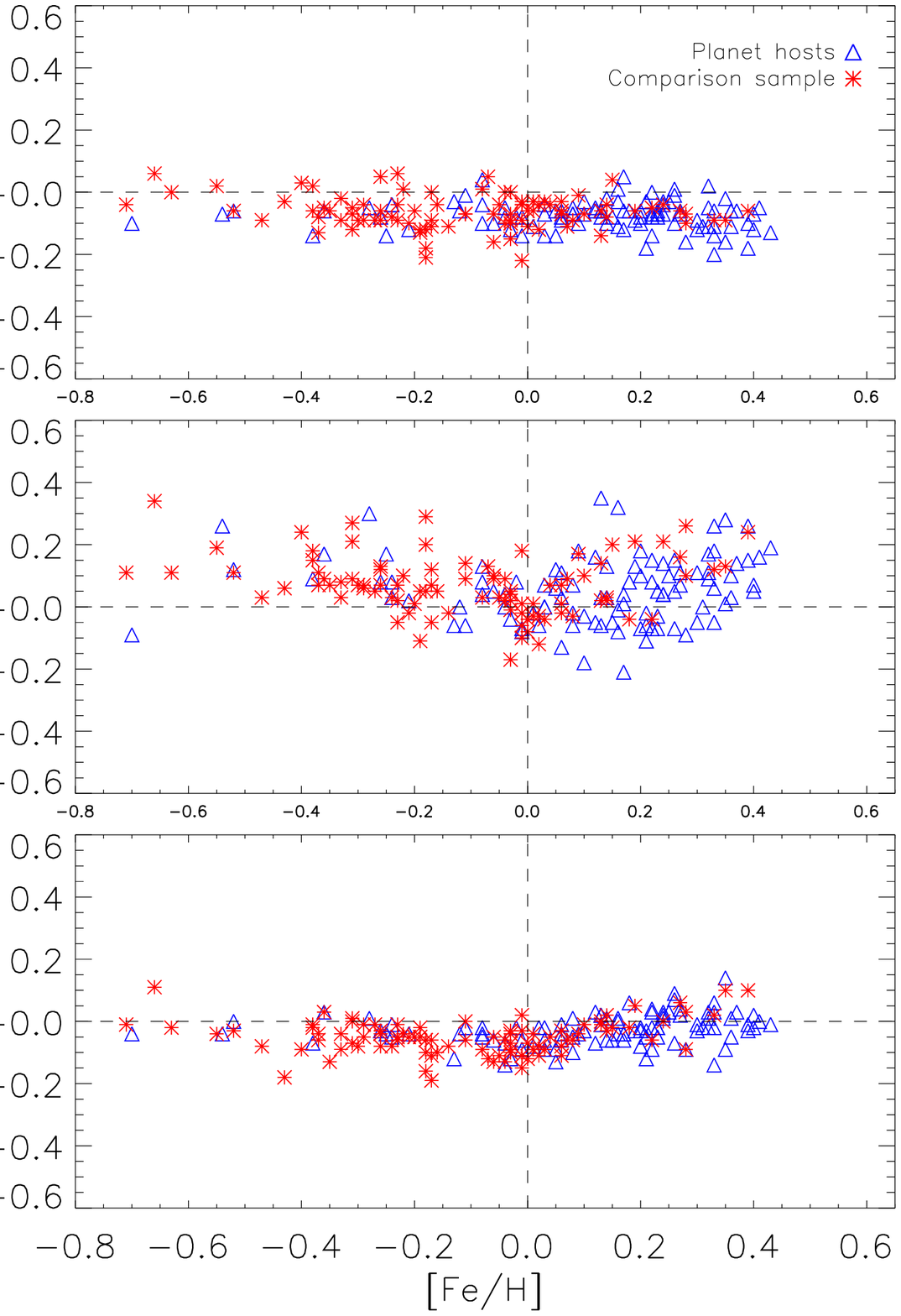}\\
\psfig{width=0.45\hsize, file=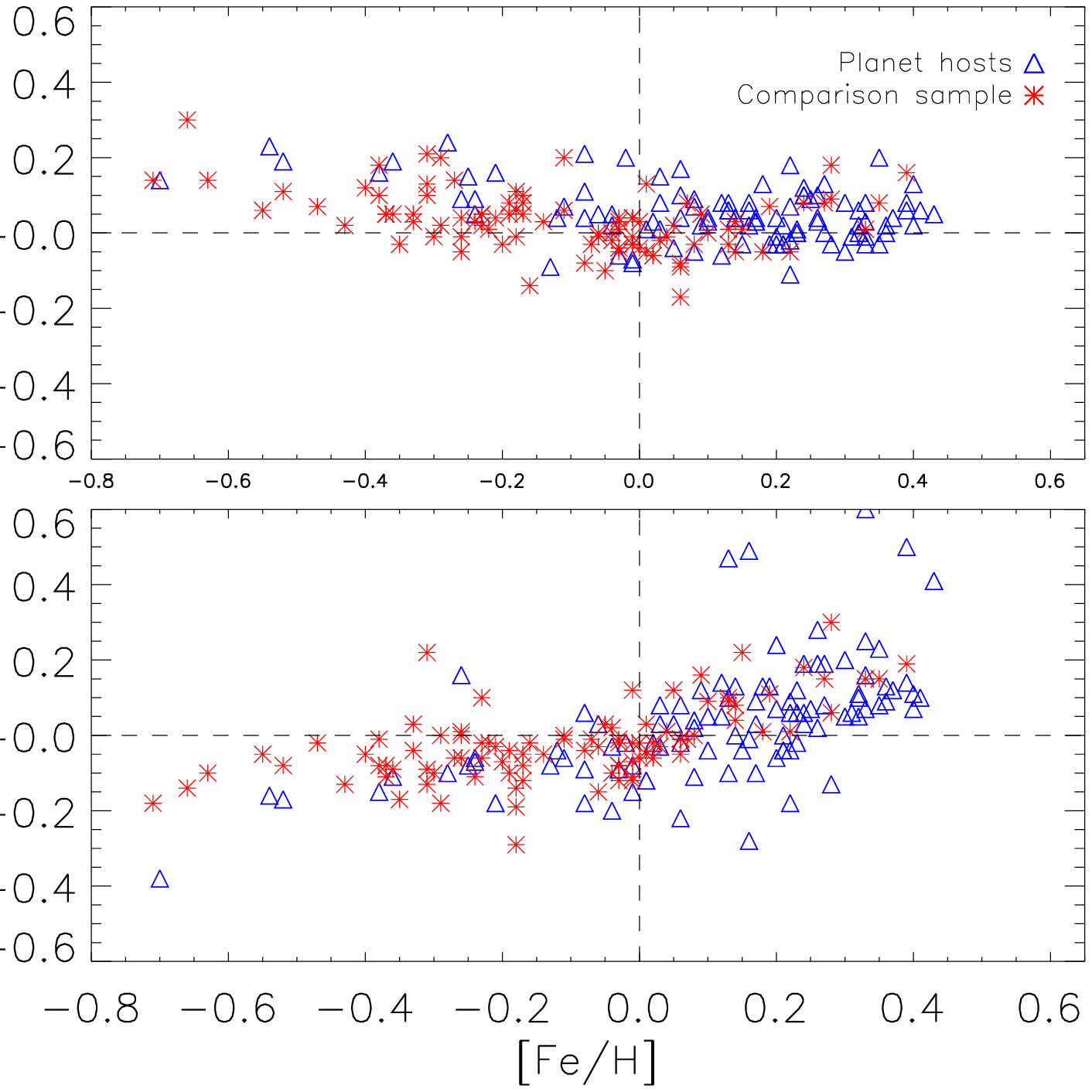}&\psfig{width=0.45\hsize, file=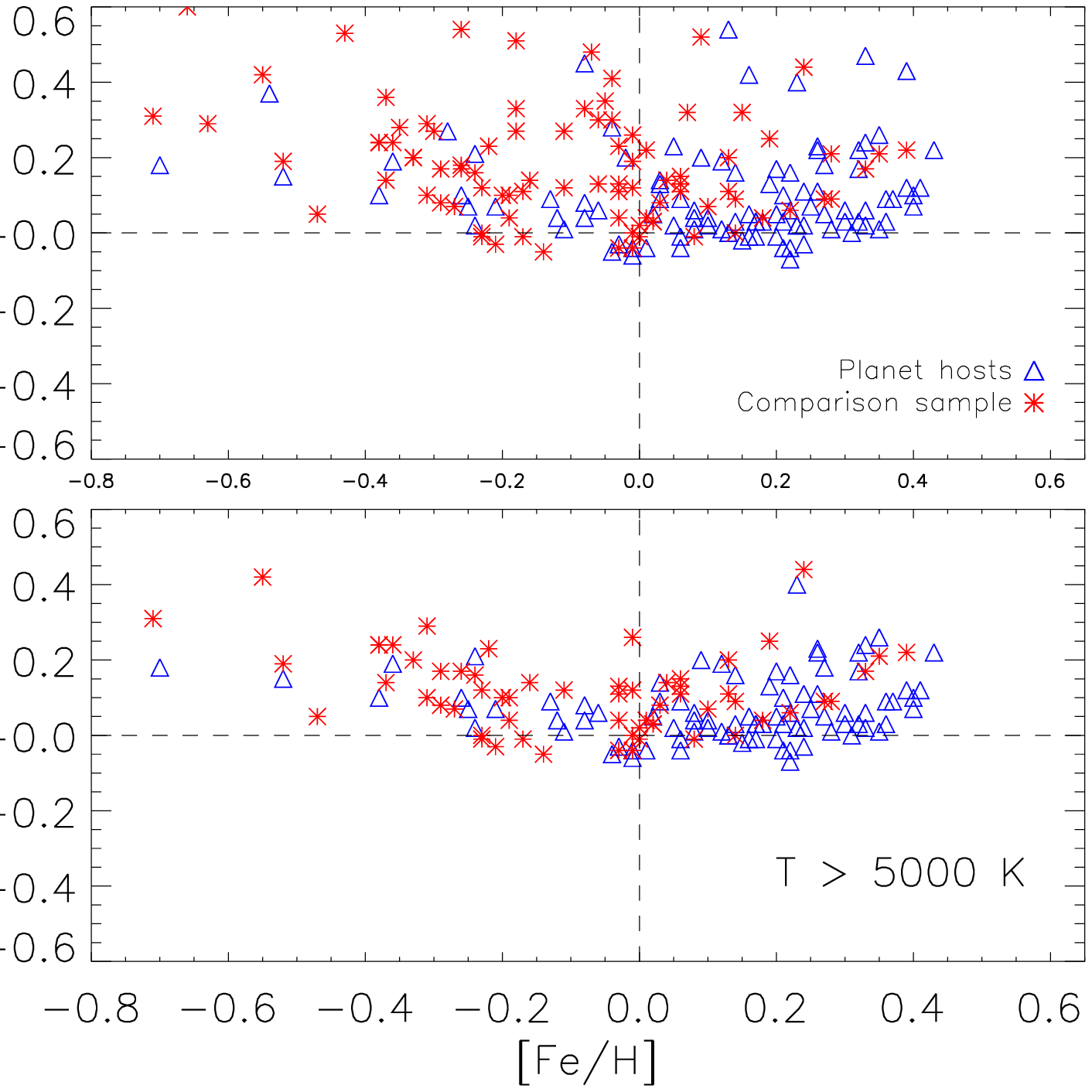}\\
\end{tabular}
\caption{$[$X$/$Fe$]$ {vs.}\ [$Fe$/$H$](in $dex$) for \textit{$\alpha$ elements} (Si, Ca, Ti, Sc) and for \textit{
iron-peak elements}  (Cr, Co, Ni, Mn, V). [V/Fe] {vs.}\ [Fe/H] for $T_{\rm eff} > 5000$ K 
objects only (below right panel). The triangles and asterisks represent stars with planets and ``single''  stars, respectively. The intersection of the dotted lines indicates the solar value.}
\label{fig1}
\end{figure*}

The histograms provide the distributions of [$X$/H] with $X$ = Si, Ca, Sc, Ti, V, Cr, Mn, Co and Ni (Fig.~\ref{isto}), and with $X$ = Na, 
Mg, Al (Fig.~\ref{istonew}) for the two samples of stars, with and without planets. These
results are similar to those presented for [Fe/H] by Santos et al.~(2004a, 2005) and clearly confirm that the observed 
metallicity excess is, 
as expected when extended to elements other than iron. We observe similar features to those noted in the [$X$/H] distribution for 
refractory elements (Bodaghee et al.~ 2003)
and for iron (Santos et al.~2001a, 2004a; Reid 2002). For example, these histograms of planet-host stars are usually not
symmetrical. This interesting feature is particularly evident for Ca, Sc, Co, Ni, Na and Al, for which the distribution seems to be 
an increasing function of [$X$/H] up to a certain
value, after which it falls abruptly. 
This cut-off corresponds to [$X$/H] $\sim 0.5$ for Si, Ti, Ni, Na, [$X$/H] $\sim 0.7$ for V, Cr, Mn and Co and 
[$X$/H] $\sim 0.6$ for Mg and Al. Only in the case of Ca does the distribution fall to [$X$/H] $\sim 0.3$. For some elements (e.g.\ Ti, V, Cr, Mg) the distributions appear to be slightly bimodal.
This is  probably related to a lack of stars with [$X$/H] $\sim 0.3$ in the planet-host sample and [$X$/H] $\sim -0.2$
 in the comparison
group for these elements.
The average values $<$[$X$/H]$>$, the rms dispersions for the two distribution, and the
difference between the average [$X$/H] for stars with and without planets are listed in Table~\ref{tabmed}. We note that this 
difference varies from 0.13 dex
for V to 0.29 dex for Mn and Na. The difference between the average abundance values of the two groups (see Table~\ref{tabmed}) 
is only an estimate and is not very significant, given the usually high dispersion around the mean values.

In Fig.~\ref{fig0} are shown [$X$/H] {vs.}\ [Fe/H] plots for refractory elements. The [$X$/H] ratio is a linear 
function of [Fe/H] and the small amount of scatter give a certain plausibility to our
results. We refer the reader to the Fig.~5 by Beir{\~a}o et al.~(2005) for the plots of [$X$/H] {vs.}\ [Fe/H] (X= Na, Mg and Al).
These plots are not presented here because adding 53 new comparison sample stars did not change the results for these elements. However,
[$X$/Fe] {vs.}\ [Fe/H] plots for Na, Mg and Al are shown in Figure~\ref{fignew}.

\section{Galactic chemical evolution trends}
With the exception of the lightest elements (e.g.\ H and He), 
the history of the Galaxy's chemical composition is dominated by nucleosynthesis occurring in many generations of stars
(McWilliam 1997). The low-mass stars are like ``fossils'' because their lifetimes are sometimes comparable to 
the age
of the Galaxy. It might actually be supposed that, at least for F--G dwarfs, the external envelope of the stars have preserved much of 
their original chemical
composition, since it has not been convectively mixed with  internal matter.
The relation [$X$/Fe] {vs.}\ [Fe/H] is traditionally used in observational studies of the chemical evolution of the Galaxy since 
iron is a
good chronological indicator of nucleosynthesis (controversial). Furthermore,  iron lines are numerous and easy to measure in
the spectra of dwarfs.

In Fig.~\ref{fignew} we present [$X$/Fe] {vs.}\ [Fe/H] plots for Na, Mg and Al taken from the work by Beir{\~a}o et al.~(2005) with the addition 
of
 abundances from a new comparison sample. 
We also did calculations relative to the iron abundances ([$X$/Fe]) of $\alpha$ elements (e.g.\ Si, Ca, Ti) and of 
iron-peak elements (e.g.\ 
Cr, Mn, Co, Ni). The former are believed to be mostly produced in the aftermath of explosions of type II supernovae  (SNe II), although, following some models,
 these elements 
might also be produced  during a
 type Ia supernova (SN Ia) event (Thielmann et al.~2002). Meanwhile, most of the latter
 would have been synthesized by SNe Ia explosions. Magnesium is supposed to be produced
by SNe II, thus comparing Mg abundances with those found for other $\alpha$ elements could provide us with  evidence concerning the 
origin
of these elements. Sodium and aluminium are thought to be mostly a product of Ne and C burning in massive stars.
It is not clear how Sc is formed, because in the periodic table it is  intermediate  between $\alpha$ and iron-peak elements. 
The origin of manganese is also debated.
However, Sc and Mn abundances in long-lived F and G stars are of great interest since they could also introduce  special constraints 
on
nucleosynthesis theory (Nissen et al.~2000). 
Even though the main aim of this study was to compare the abundances of refractory elements in stars with and without planets,
 our results also 
give us a chance to increase our
present knowledge of the chemical evolution of the Galaxy at [Fe/H] $>0$. 
 In Figs~\ref{fig1} and~\ref{fignew} we present the [$X$/Fe] {vs.}\ [Fe/H] plots for all the elements.

\section{Comparison with the literature}
There are already several studies concerning the chemical abundances of elements other than iron in F, G, K main sequence stars in the solar
neighborhood (see Edvardson et al.~1993; Feltzing \& Gustafsson 1998; Chen et al.~2000; Nissen et al.~2000; Fulbright et al.~2002; Reddy
et al.~2003; Bensby et al.~2003; Allende Prieto et al.~2004). In addition, various studies on chemical abundances in
planet-host stars have gradually emerged (Gonzalez et al.~2000; Santos et al.~2000; Takeda et al.~2001; Sadakane et al.~2002; Bodaghee 
et al.~2003; Ecuvillon et
al.~2004a, 2004b; Beir{\~a}o et al.~2005; Fischer \& Valenti 2005).
We note that our abundance trends generally agree with those published in the literature and this lends a certain reliability to our
 results. 
 In the following subsections, we  describe [$X$/Fe] {vs.}\ [Fe/H] trends for each element and then  make a brief
comparison with studies in the literature on this subject. This comparison is divided into two parts:  first, 
we compare studies on abundances in stars with planets; second,  we extend the comparison to other studies regarding
abundance trends in metal-rich stars of the Galactic disc. We  focus   our attention mainly on the [Fe/H] $>0$ range.
Given the unobserved or probably insignificant differences
between the two samples of stars presented, we  consider the distribution as a whole for the rest of the analysis. 
Possible differences between our trends and those recently published could be of great interest since it might reflect the presence of
planets. In contrast, any relation  to the presence of a planet is perhaps coincidental and the global trends observed 
are probably best interpreted as a consequence of  Galactic chemical evolution.

\subsection{The $\alpha$ elements}

\subsubsection{Silicon}
\begin{figure*}[!hbt]
\begin{tabular}{lcr}
\psfig{width=0.30\hsize, file=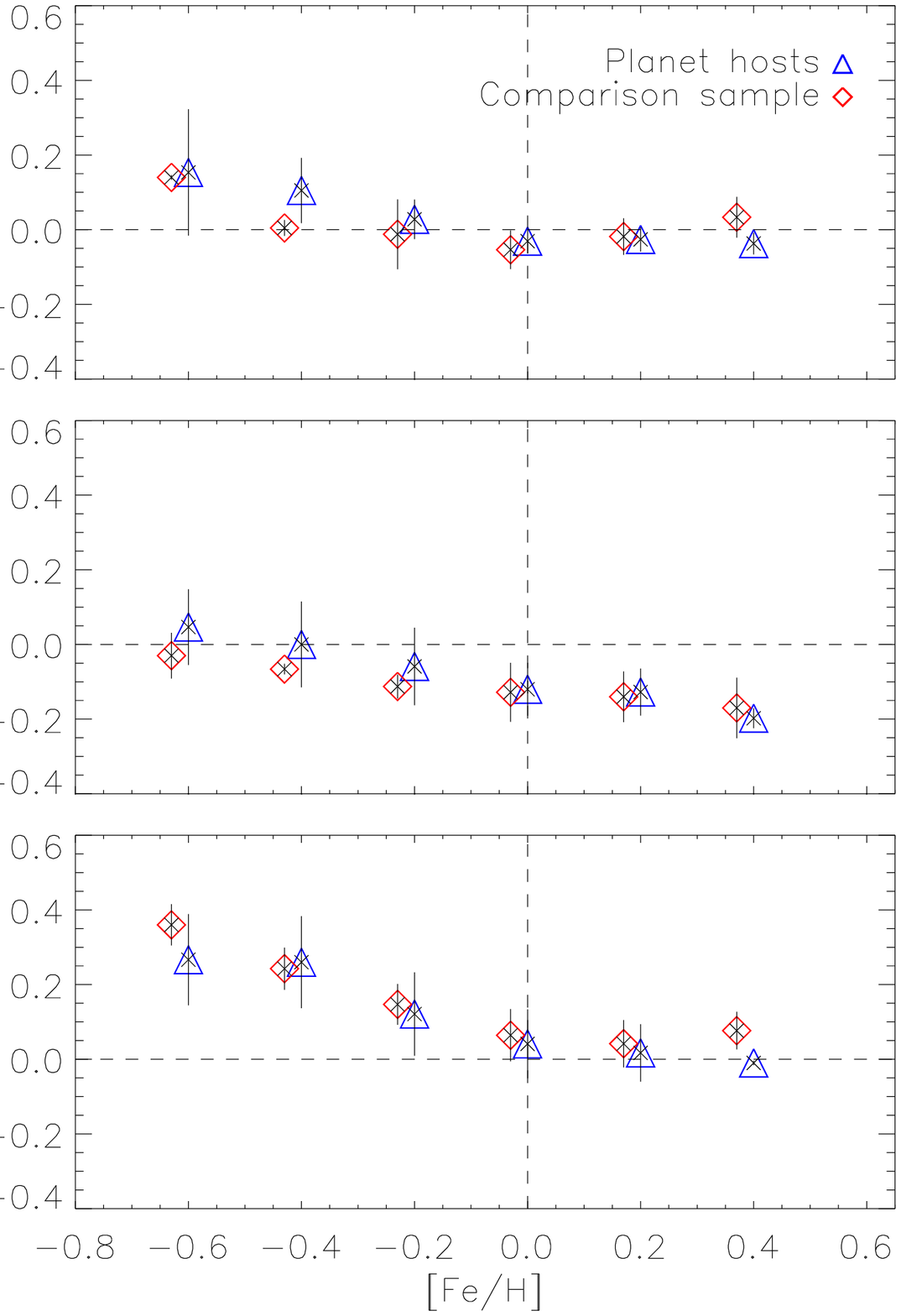}&\psfig{width=0.30\hsize, file=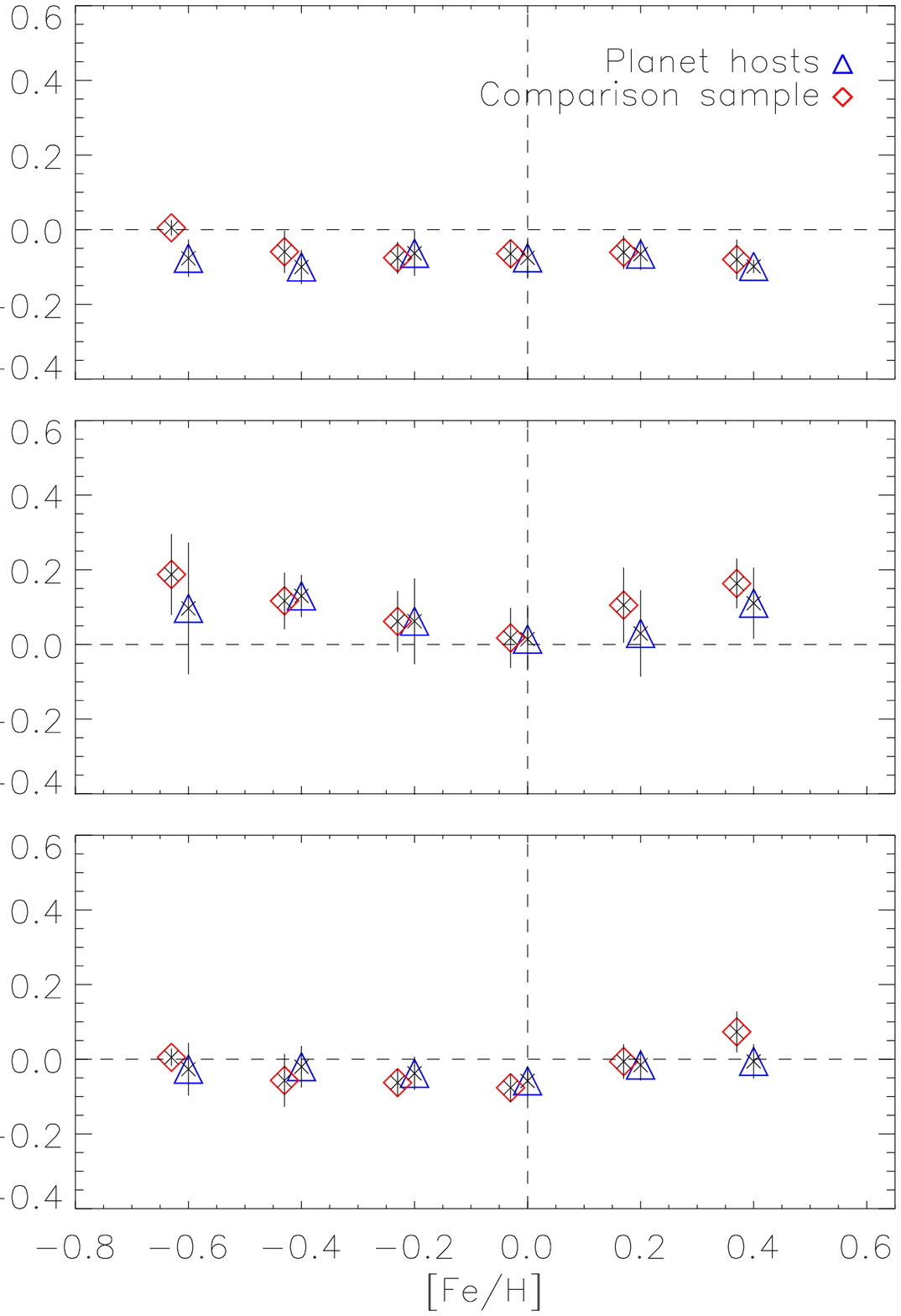} &\psfig{width=0.30\hsize,file=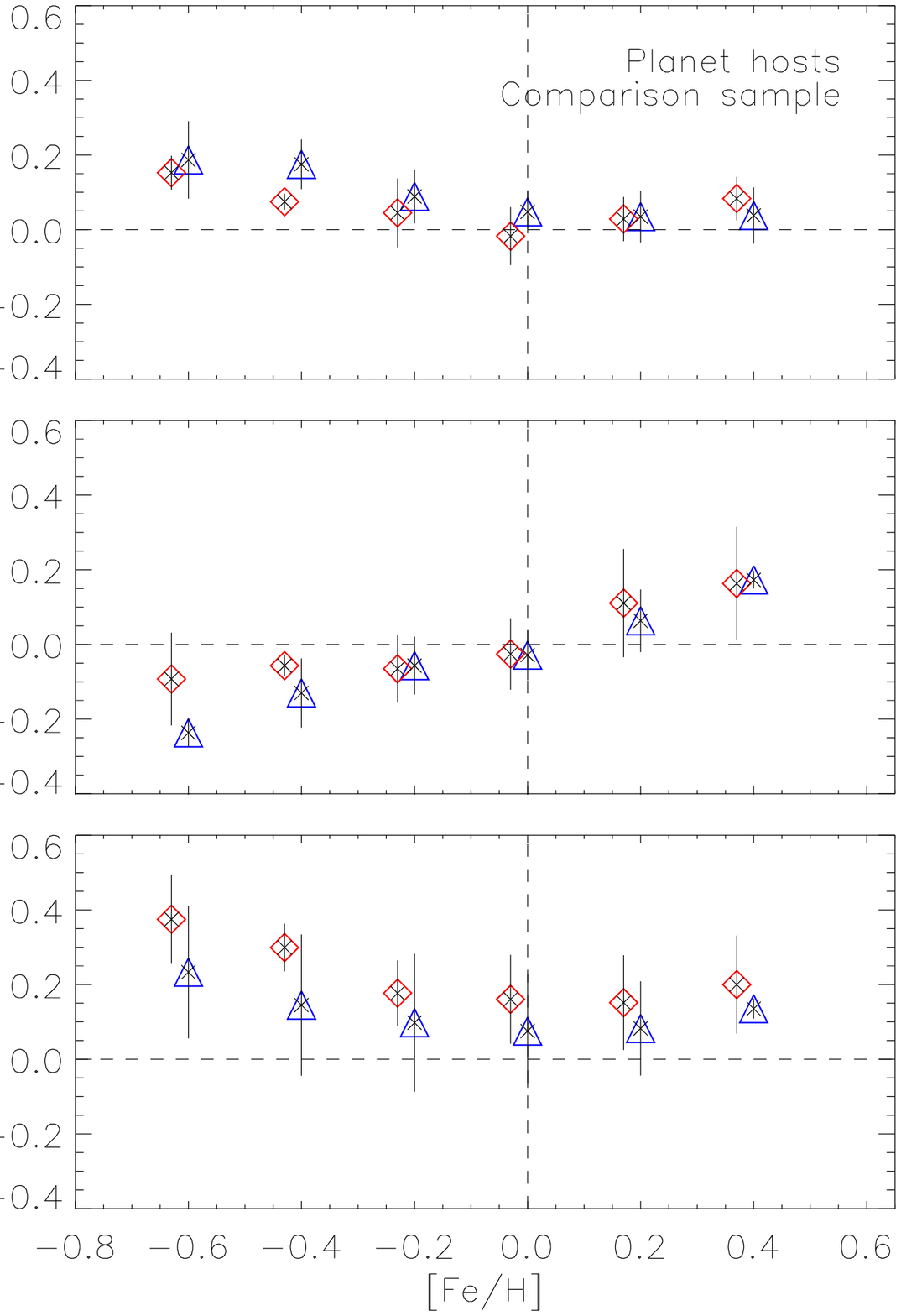}\\
\end{tabular}
\caption{[$X$/Fe] {vs.}\ [Fe/H] trends using binned average values. The bins are 0.2 dex wide, centred on [Fe/H] $=-0.6,
-0.4,-0.2,0,0.2$ and 0.4
for both samples. Triangles are planet-hosts while rhombi are stars without a planetary-mass companion. The error bars represent the
standard deviation about the mean value.}
\label{andam}
\end{figure*}

\begin{figure*}[tb]
\centering
\psfig{width=0.9\hsize,file=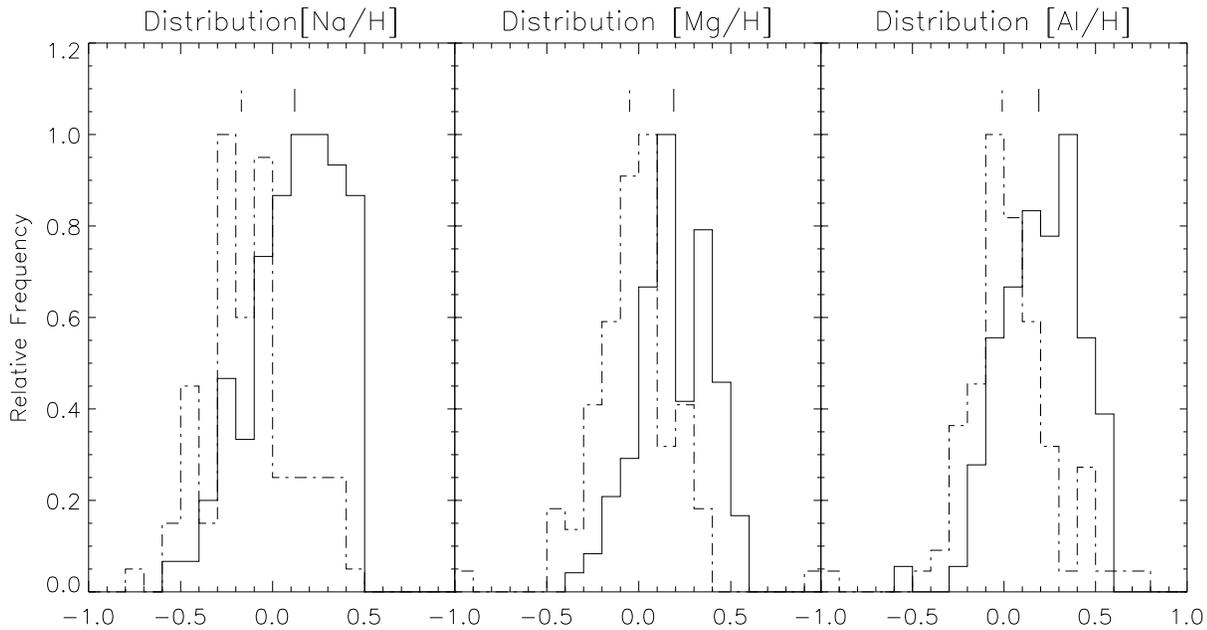}
\caption{Same as Fig.~\ref{isto} but for the Na, Mg and Al distributions.}
\label{istonew}
\end{figure*}

In Figs~\ref{fig1} and~\ref{andam} we note a slight [Si/Fe] overabundance compared to the solar value at 
[Fe/H] $\lesssim -0.2$ (with a rather larger scatter), while the point distribution remains constant at [Si/Fe] $\sim 0.0$ for the rest of the metallicity range.

Our results agree with the previous chemical analyses of stars with planets by Gonzalez et al.~(2000, hereafter GZ00), Bodaghee et
al.~(2003, hereafter BOD) and Fischer \& Valenti (2005). Sadakane et al.~(2002, hereafter SD02) in their study analysed only a few objects (12 stars with extrasolar planets) but their abundance 
distribution is flat at high metallicities.

Edvardsson et al.\ (1993, hereafter EAGLNT) and Chen et al.\ (2000, hereafter C00) obtained similar results, as well as 
Bensby at al.\
(2003, hereafter BEN) and Fulbright et al.\ (2002, hereafter FUL) in the range [Fe/H] $>0$. In another study on
metal-rich stars ([Fe/H] $>0.10$) (Felzing \& Gustafsson 1998, hereafter FG98) the [Si/Fe] distribution exhibits a constant 
trend around the solar value. All these results differ from
those of Allende Prieto et al.\ (2004, hereafter AL04), in which the abundance trend for Si changes abruptly around [Fe/H] $\sim 0$ 
and assumes a
positive slope.

\subsubsection{Calcium}
Contrary to the other $\alpha$ elements, the  [Ca/Fe] ratio seems to decrease quite uniformly (see Fig.~\ref{fig1} 
and Fig.~\ref{andam}).
In particular, the plot suggests the presence of a {plateau} in the range $-0.2\lesssim$ [Fe/H] $ \lesssim 0.2$  
followed, for higher
metallicities, by a slight fall-off. Planet-host stars with $T_{\rm eff}<$ 5000 K show a certain dispersion in [Ca/Fe] 
values, for example the stars HD~177830, 
HD~137759 and HD~114783.

The similar behaviour of calcium trends for  stars both with and without planets has been observed in other studies of chemical abundances 
(BOD and GZ00). As BOD have noticed,  SD02 analysed only two stars in the range $0.2\lesssim$ [Fe/H] $\lesssim 0.4$ 
(where a possible {plateau} is suggested), so the comparison with our results is only partially valid.

The Ca distribution appears quite ``flat'' in the data presented by EAGLNT, FG89, AL04 and FUL. However, the results plotted by
BEN appear to decrease for [Fe/H] $\sim 0.2- 0.3$.

\subsubsection{Titanium}
The titanium  [Ti/Fe] ratio decreases in the range  $-0.6\lesssim[Fe/H]\lesssim 0$ until the solar value where the
distribution settles (Fig.~\ref{fig1} and~\ref{andam}). 

Similar trends have been obtained by BOD and Fischer \& Valenti (2005), while we cannot observe any slope change around 
$-0.2\lesssim$ [Fe/H] $\lesssim 0$ in SD02, because only a few objects have been plotted. Despite the large scatter in the 
GZ00 data, titanium abundances seem to decrease gradually with metallicity.

A quite pronounced point dispersion  is also observed in EAGLNT, BEN and CH00, especially for [Fe/H] $\lesssim -0.3$ (as seen here).
This scatter is probably due to the overestimation of EWs caused by the increasing blending effect (which becomes severer as $T_{\rm eff}$
falls) or by Galactic evolution effects. Finally, comparing the trends for [Fe/H] $>0$ we note that [Ti/Fe] values
remain approximately constant in the plots of  BEN, FUL and FG98, 
while the data of AL04 again show a rise above solar metalicity.

\subsubsection{Scandium}
Figures~\ref{fig1} and~\ref{andam} illustrate that the scandium trend is similar to that of other $\alpha$ elements. It drops until
$-0.2\lesssim$ [Fe/H] $\lesssim 0$ and afterwards the [Sc/Fe] ratio approximately  follows the solar value at high metallicities. 

Previous studies on stars with planets have obtained similar results, as presented in BOD and SD02.

\begin{figure*}[]
]\begin{tabular}{lr}
\psfig{width=0.45\hsize, file=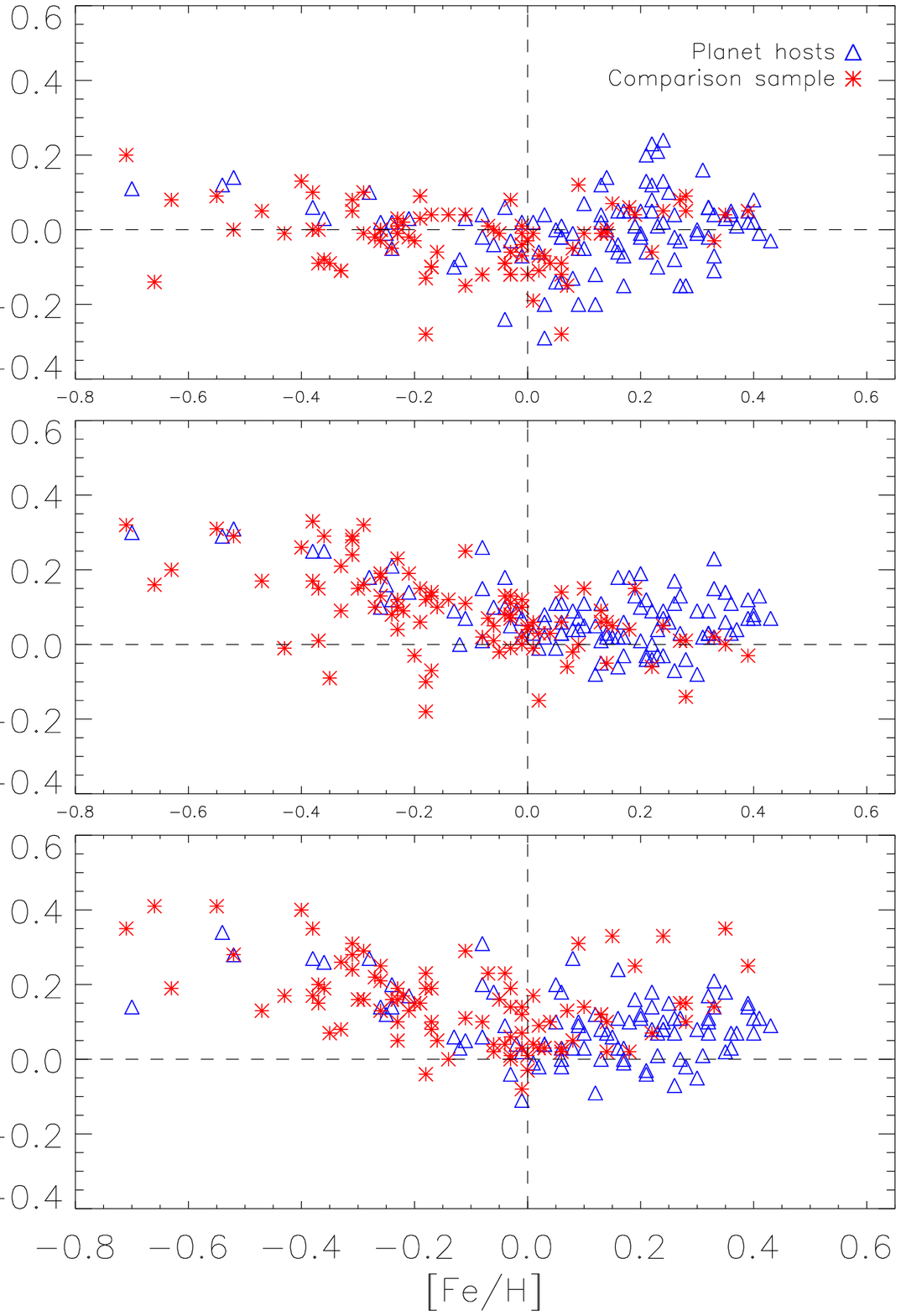}&\psfig{width=0.45\hsize, file=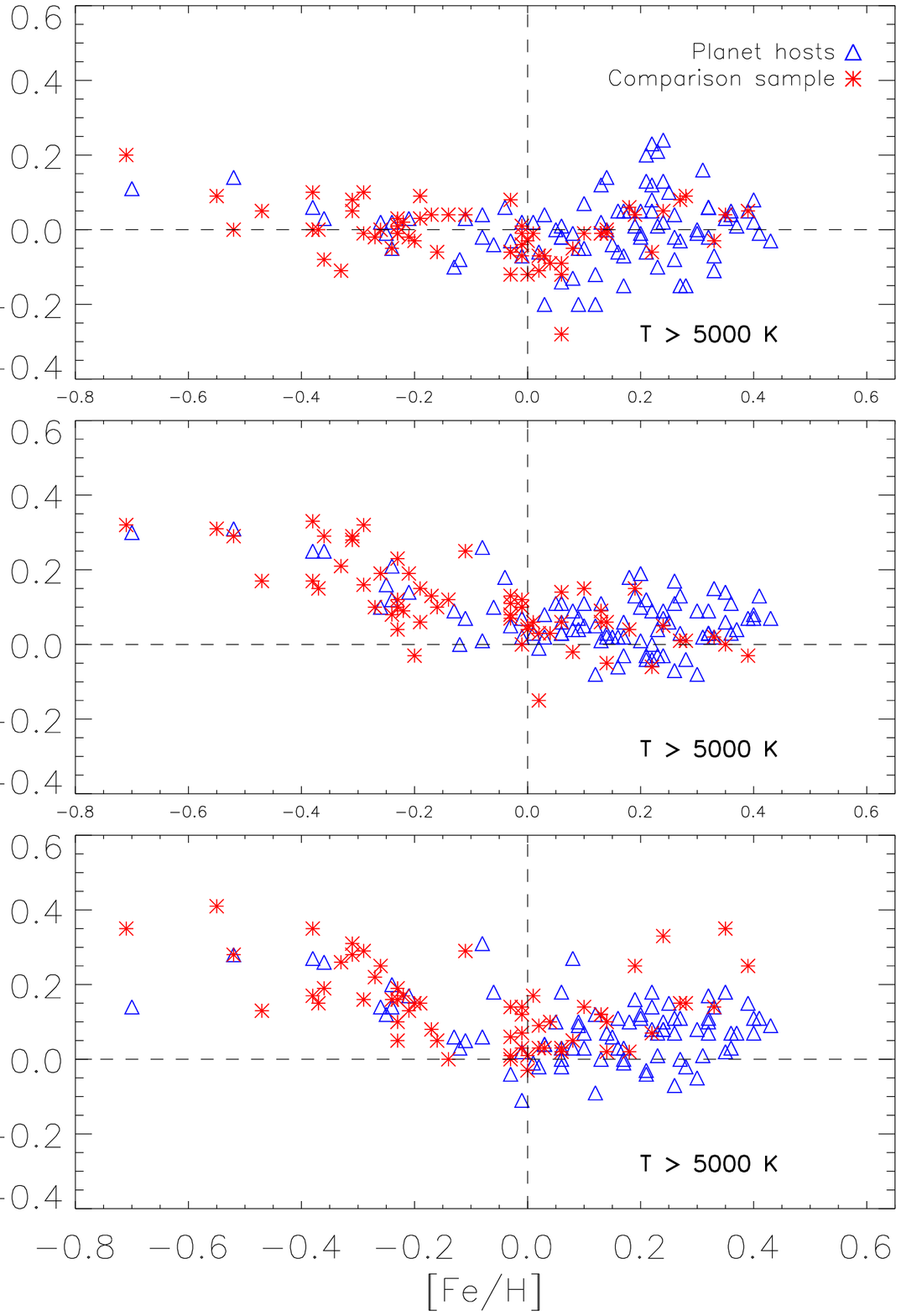}\\
\end{tabular}
\caption{[$X$/Fe] {vs.}\ [Fe/H] (in dex). Results by Beir{\~a}o et al.\ (2005) with the addition of abundances from new 
comparison
sample spectra. On the left are the plots for Na, Mg and Al, and on the right
the same plots only for $T_{\rm eff}>5000$ K objects.
 Triangles and asterisks represent stars with planetary-mass
companions and ``single'' field stars, respectively. The intersection of the dotted lines indicates the solar value.}
\label{fignew}
\end{figure*}

There are only a few studies in the literature about scandium abundances. The most detailed one was presented by Nissen et al.\ (2000) in the
range $-1.4<$ [Fe/H] $<0.1$, so comparison with our results at high metallicity is not possible at all. However, it is 
interesting to note a slope
change at [Fe/H] $\sim -0.3, -0.2$ that is also represented in our graphs. Results by FG98 show a substantial star-to-star scatter
but globally the [Sc/Fe] ratio remains around solar for [Fe/H] $>0$. In contrast, the AL04 plot shows  large enrichments 
in scandium
for metal-rich stars.

\subsection{The Fe-group elements}

\subsubsection{Manganese}
Manganese is one of the lesser studied element in the literature.
Manganese lines turn out to be the most difficult  to measure
 owing to unknown blended lines that probably cause the 
quite large
point spread, particularly for stars with planets. 
The [Mn/Fe] ratio generally tends to increase  with the metallicity, %  expecially for metal rich stars,
 differently from other iron-peak elements (see Fig.~\ref{andam}). In Fig.~\ref{fig1} we observe a clear change of slope around 
 $-0.2\lesssim$ [Fe/H] $\lesssim 0$.

Despite a certain scatter, we note the good agreement between our results and those of BOD and SD02. 

Nissen et al.~(2000) analysed Mn abundances, in disc stars ($-1.4<$ [Fe/H] $<0.1$). These results show an evident increase in
 manganese abundances for
 the entire metallicity range considered. Another study (FG98) also exhibits a slight linear [Mn/Fe]  dependence on iron in 
metal-rich stars.

\subsubsection{Vanadium}
On the subject of vanadium abundances, Fig.~\ref{fig1} (bottom right) clearly shows  how removing cooler stars ($T_{\rm eff} < 5000$~K) from the 
data considerably reduces the dispersion of points. 
Although vanadium belongs to  the iron-peak group, we note that [V/Fe] ratio behaves like an $\alpha$ element. Figure~\ref{andam} shows that [V/Fe] values in 
stars with planets are systematically $\sim 0.10$ dex lower than in the comparison sample stars.
Since planet-hosts are, on average, hotter than comparison sample stars (see Fig.~\ref{istemp}) and [V/Fe] shows a negative slope with
$T_{Â\rm eff}$, this effect may contribute to the observed difference.

The same considerable scatter is observed in the  studies by BOD and SD02, but good agreement among trends is found in both
cases. Contrary to our results and those of BOD, SD02 emphasized that the [V/Fe] values of stars with planets are about 0.15 dex 
higher than comparison
sample vanadium abundances. This mismatch could suggest that vanadium analysis is strongly influenced by an NLTE effect, as discussed in
Section~\ref{erro}.

There are not many studies regarding vanadium abundances in metal-rich stars without planets. The [V/Fe] ratio remains around
the solar value in the plots of FG98 and CH00, for [Fe/H] $>0$.

\begin{figure}[bt]
\begin{tabular}{r}
\psfig{width=0.90\hsize, file=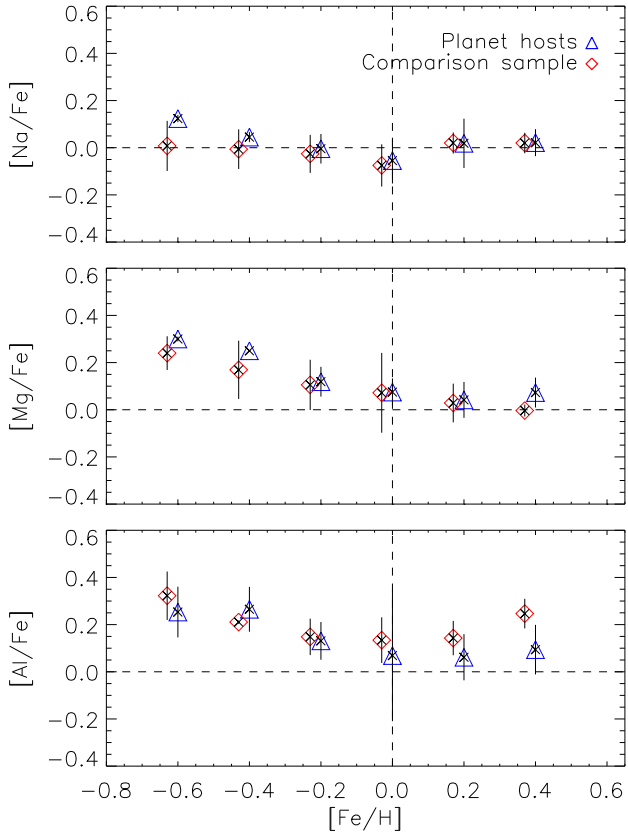}\\
\end{tabular}
\caption{The same as Fig.~\ref{andam} but for Na, Mg and Al.}
\label{andan}
\end{figure}

\subsubsection{Chromium}
Chromium abundances in our target are constant around [Cr/Fe] $\sim -0.05$ dex (see Fig.~\ref{andam}). We note  very little
 scatter for this element (Fig.~\ref{fig1}).

Our observed trend is similar to those already published by BOD and SD02: the point distribution is uniform, constant and with  
little scatter. However, the BOD data show about 0.05 dex systematically lower chromium abundances compared to our results.

Good agreement is found with the FG98, CH00, BEN and FUL plots. All these results exhibit a [Cr/Fe] ratio of around 0.

\subsubsection{Cobalt}
The [Co/Fe] ratio first decreases to the solar value and then slowly rises for metal-rich stars (see also Fig.~\ref{andam}). The figures also show that
the abundances of metal-rich stars with planets seems to be lower ($\sim 0.05$ dex) than [Co/Fe] values for stars without planets.

Previous studies of BOD and SD02, have obtained similar results for cobalt. In this last
paper authors suggested that planet-host stars exhibit a slight Co overabundance with respect to the solar value ($\sim
0.15$ dex) when compared to  stars with no companions.

Our results resemble those of FG98, while with respect to AL04 we obtained a change of  slope of around $[Fe/H]\sim 0$.

\subsubsection{Nickel}

 Similarly to calcium and vanadium, the nickel trend shows a possible {plateau} in the range  $-0.2\lesssim$ [Fe/H] $\lesssim 0$ and hence a slight slope 
change for metal-rich stars (see Fig.~\ref{fig1}).

Our trends resemble those observed by BOD, SD02 and Fischer \& Valenti (2005). The [Ni/Fe] ratio seems to increase slightly in 
metal-rich
stars.

In most of the studies considered (see FG98, CH00, FUL, EAGLNT) nickel abundances exhibit a uniform and approximately constant
trend around the solar value. In  the results proposed by BEN, nickel abundances remain constant until [Fe/H] $\sim 0.2$, where the 
values
are greater than the rest of the set of data.
  A different plot is represented by AL04: the [Ni/Fe] ratio increases
for higher metallicity values.

\subsection{Na, Mg and Al}

In Figs~\ref{fignew}-\ref{andan} we present the plots for Na, Mg and Al. As discussed in detail in the paper by Beir{\~a}o et 
al.~(2005), the [Na/Fe] 
ratio slowly decreases as a function of [Fe/H] until the solar metallicity, then we can observe a change of the slope. This 
behaviour is similar to that
shown by some refractory abundance distribution (e.g.\ Si, Ti, Sc). Despite the dispersion of the points, the [Na/Fe] values are on
average below solar for [Fe/H] $\sim$ 0.0.  

The Mg and Al abundances also resemble those derived for Si, Ti, Sc. We note that the decrease in the [Mg/Fe] and [Al/Fe] values with
increasing metallicity in the range $-0.70<$ [Fe/H] $< 0$ is stronger than for Na. For metal-rich stars the abundance distributions
 for the three
elements stay at approximately  solar value  except for a light upturn in [Na/Fe] and [Al/Fe] values. 

These results  agree globally with previous studies of stars with planets (see SD02, GZ00 and Fischer \& Valenti 2005 for Na plots) and 
we also
confirm that Mg and Al may be slightly enhanced in the planet-host HD 168746, as noticed in SD02. The [Na/Fe] upturn above solar
metallicity is also found in GZ00 and both EAGLNT, FG98, BEN, but not in CH00. With respect to Al abundances, other authors (EAGLNT, 
FG98,
BEN) have observed similar trends. In the plots proposed by CH00 we note a clear upturn at [Fe/H] $\sim -0.2,-0.3$, as traced by 
comparison stars in our sample.

Possible differences between the abundance trends of the two groups of stars  have been observed here only for metal-rich stars. 
It is particularly interesting to note that comparison stars continue to show a decrease in [Mg/Fe] with increasing [Fe/H], while maintaining the 
slope observed for lower metallicity, while planet-hosts change the slope. An opposite effect is observed for [Al/Fe] where
the planet hosts seem to have less Al than their ``single'' conterparts. The dependence of [Mg/Fe] on
$T_{\rm eff}$ is not responsible for this effect (see figs 1 and 2 in Beir{\~a}o et al.~2005) and we cannot rule out 
that this difference is real. 
Figure~\ref{fignew} also shows that removing cooler stars ($T_{\rm eff}< 5000 K$) 
from the dataset does not change these results. GZ00 also suggested a difference in [Mg/Fe] values between planet-host
and single stars.

\section{Discussion and conclusion}

We have determined abundances for nine refractory elements (other than iron) in a large sample of 101 stars with planets and in
a homogeneous comparison sample of 94 stars with no known planets. We have also presented Na, Mg, Al abundances in 53 new comparison
sample stars to extend the previous work by Beir{\~a}o et al. (2005). For each element  a uniform and independent study 
of the two samples was carried out
using atmospheric parameters derived from a detailed spectroscopic analysis by Santos et al.\ (2004a, 2005).
Abundance ratios [$X$/H] {vs.}\ [Fe/H] and [$X$/Fe] {vs.}\ [Fe/H] have been plotted to compare the two samples 
and to try to find differences eventually
connected to the presence of giant planets. This study was also intended to provide a complete comparison in the  high metallicity domain, 
where
studies had lacked ``single'' stars with [Fe/H]$>0.1$.
Furthermore, the data could provide clues clarifying the chemical evolution of planetary systems.

In our analysis we stressed a certain diversity of trends for elements of a common origin.  On one hand not all the refractory elements 
studied here show the same behaviour;  on the other hand, the abundance trends of elements coming from the same nucleosynthesis source are not
always alike, in contrast with that we expected.

Our concluding remarks are as follows:
\begin{itemize}
\item{Again we confirm that the excess of metallicity observed for planet hosts is not unique to iron.}
\item{The abundance trends of stars with planets are very similar to those traced by comparison sample stars. This feature could
favour the primordial hypothesis to explain the metallicity excess in stars harbouring planets. In any case, some elements (e.g.\ Mg, Al, V, 
Co) show certain differences in
the behaviour of abundances of stars with planets and ``single'' stars, in the higher metallicity range. We thus do not exclude the possibility that the
presence of a planet might influence the composition of certain elements in the atmosphere of  metal rich stars.}
\item{Good agreement was found with both previous published studies on abundances in stars with extrasolar planets and most studies
on metal-rich stars of the Galactic disc. One might suggest that the observed trends are simply a conseguence of  Galactic chemical 
evolution,
with no particular mechanism linked to the presence of a planet. To this end, only the calcium abundances show a different trend 
when comparing studies on stars with planets (BOD, GZ00 and this article) and most chemical analysis on Galactic disc stars (EAGLNT, FG98,
AL04, FUL).}

\end{itemize}

In the future  gathering new homogeneous abundance values of other elements  with a wide range of
condensation temperatures $T_{\rm C}$ will be of great importance.
For example, a detailed comparison of [$X$/Fe] abundances of volatile and refractory elements
is currently in progress . This study will 
give us the chance to discuss the relative importance of
differential accretion (e.g.\ Gonzalez 1997; Smith et al.~2001; Sadakane et al.~2003) in stars harbouring extrasolar planets.

\begin{acknowledgements}
 Support from Funda\c{c}\~ao para a Ci\^encia e a Tecnologia (Portugal)
to NCS in the form of a scholarship 
(reference  SFRH/BPD/8116/2002) and a grant (reference 
POCI/CTE-AST/56453/2004) are gratefully acknowledged. Thanks to the anonymous
referee for his/her useful suggestions.
 \end{acknowledgements}

%---------------------------bibliography---------------------------


\begin{thebibliography}{}

\bibitem{ALL}  Allende Prieto, C., Barklem, P.~S., Lambert, D.~L. \& Cunha, K. 2004 A\&A 420, 183

\bibitem{AG} Anders. E., \& Grevesse, N.~1989, Geochim, Cosmochim. Acta, 53, 197

\bibitem{Bei} Beir{\~a}o, P., Santos, N.~C., Israelian, G. \& Mayor, M.~2005, A\&A 438,
251


\bibitem{BEN} Bensby, T., Feltzing, S. \& Lundstr\"{o}m, I. 2003, A\&A, 410, 527 (BEN)
\bibitem{BEN2} Bensby, T., \& Ilyn, I., 2005 A\&A, in press 

\bibitem{BODAG} Bodaghee, A., Santos, N.~C., Israelian, G. \& Mayor, M.~2003, A\&A, 404, 717 (BOD)


\bibitem{DEL} Deliyannis, C.~P., Cunha, K., King, J.~R., \& Boesgaard, A.~M. 2000, AJ, 119, 2437

\bibitem{DES} Desidera, S., Gratton, R.~G., Endl, M., Claudi, R.~U., \& Cosentino, R. 2004, A\&A, 420, L27 

\bibitem{CH} Chen, Y., Q., Nissen, P.~E., Zhao, G., Zhang, H.~W., \& Benoni, T.~2000, A\&AS, 141, 491 (CH00)



\bibitem{EC1} Ecuvillon, A., Israelian, G., Santos, N.~C., et al. 2004a, A\&A, 418, 703
\bibitem{EC2}  Ecuvillon, A., Israelian, G., Santos, N.~C., et al. 2004b, A\&A, 426, 619
\bibitem{EC3}  Ecuvillon, A., Israelian, G., Santos, N.~C., et al. 2005, A\&A, submitted

\bibitem{EDV} Edvardsson, B., Andersen, J., Gustafsson, B., et al.~1993, A\&A, 275, 101 (EAGLNT)  

\bibitem{FZ} Feltzing, S., \& Gustafsson, B. 1998, A\&AS, 129, 237 (FZ98)

\bibitem{fsc} Fischer, D.\& Valenti, J. 2005, ApJ, 622, 1102-1117

\bibitem{FULB} Fulbright, J.~P. 2002, AJ, 123, 404 (FUL)

\bibitem{RAM} Garc{\'i}a L{\'o}pez R., \& P{\'e}rez de Taoro, M.~R. 1998, A\&A, 334, 599

\bibitem{GRAY} Gray, D., 1992, in: ``The observation and analysis of stellar photospheres'', Cambridge Univ.~Press

\bibitem{GOZA} Gonzalez, G., 1997, MNRAS, 285,403

\bibitem{GOZ} Gonzalez, G., 1998, A\&A, 334, 221

\bibitem{GOZ0} Gonzalez, G., \& Laws, C. 2000, A\&A, 119, 390

\bibitem{GOZ1} Gonzalez, G., Laws, C., Tyagi, S., \& Reddy, B. 2001, AJ, 121, 432

\bibitem{GUV} Gurtovenko, E.~A., \& Kostyk, R.~I. 1989, Fraunoffer spectrum and system of solar oscillator strengths, KiIND, 200

\bibitem{GAR2} Israelian G., Santos, N.~C., Mayor, M., \& Rebolo, R. 2001, Nature, 411, 163 

\bibitem{GAK} Israelian G., Santos, N.~C., Mayor, M., \& Rebolo, R. 2003a, A\&A, 405, 753


\bibitem{GAR3} Israelian G., 2003b, in IAU S219: ``Stars as Sun: Activity, Evolution, and Planets, ed A.~K.~Dupree (San Francisco: ASP) 

\bibitem{GAK1} Israelian G., Santos, N.~C., Mayor, M., \& Rebolo, R. 2004, A\&A, 414, 601


\bibitem{KUR} Kurucz, R.~L., 1993, CD-ROMs, ATLAS9 Stellar Atmospheres Programs and 2 km $s^{-1}$ Grid (Cambridge: Smithsonian Astrophys.~Obs.)

\bibitem{KUR} Kurucz, R.~L., Furenlid, I., Brault, J., \& Testerman, L. 1984, Solar Flux Atlas from 296 to 1300 nm, NOAO Atlas No.~1

%\bibitem{MARC} Marcy, W.~G., \& Butler, R.~P., 1998, ARA\&A, 36, 57

\bibitem{MQ} Mayor, M., \& Queloz, D. 1995, Nature, 378, 355

\bibitem{MCW} McWilliam, A., 1997, ARA\&A, 35, 503

\bibitem{MOORE} Moore, C.~E., Minnaert, M.~G.~J, \& Houtgast, J. 1966, The solar Spectrum 2934 \AA\, to 8770 \AA

\bibitem{MUC} Murray, N., \& Chaboyer, B. 2002, 566, 442

\bibitem{LAW1} Laws, C., \& Gonzalez, G. 2001, ApJ, 553, 405
\bibitem{LAW2} Laws, C., Gonzalez, G., Walker, K.~M., et al.~2003, AJ, 125, 2664

%\bibitem{LAG} Laughlin, G. 2000, ApJ, 545, 1064

\bibitem{NIS} Nissen, P.~E., Chen, Y.~Q., Schuster, W.~J., \& Zhao, G.~2000, A\&A, 353, 722

\bibitem{PIN} Pinsonneault, M.~H., DePoy, D.~L., \& Coffee, M. 2001, AJ, 556, L59

\bibitem{REI} Reid, I.~N. 2002, PASP, 114, 306

\bibitem{RED} Reddy, B.~E., Tomkin, J., Lambert, D.~L., \& Allend Prieto, C., 2003, MNRAS, 340, 304

\bibitem{SD} Sadakane, K., Ohukubo, M., Takada, Y., et al.~2002, PASJ, 54, 911

\bibitem{SAN00} Santos, N.~C., Israelian, G., Mayor, M. 2000, A\&A, 363, 228
\bibitem{SAN3} Santos N.~C., Israelian, G., \& Mayor, M. 2001a, A\&A, 373, 1019 
%\bibitem{SAN0} Santos, N.~C, Mayor, M., Naef, D. 2001b, A\&A, 379, 999 
\bibitem{SAN6} Santos N.~C., Israelian, G., Mayor, M., 2001b in ``Confirming the Metal-Rich Nature of Stars with Giant Planets. Proceedings of the 12th Cambridge workshop``Cool Stars, Stellar System, and the Sun'', Boulder, Colorado, USA

\bibitem{SAN7} Santos, N.~C., Garc{\'i}a L{\'o}pez R.~J., Israelian, G., et al. 2002, A\&A, 386, 1028
\bibitem{SAN1} Santos, N.~C., Israelian, G., Mayor, M., Rebolo, R., \& Udry, S.~2003, A\&A, 398, 363

\bibitem{SAN2} Santos N.~C., Israelian, G., \& Mayor, M. 2004a, A\&A, 415, 1153

\bibitem{SAN2b} Santos N.~C., Israelian, G., Garc{\'i}a L{\'o}pez R., et al.~2004b, A\&A, 427, 1085 

\bibitem{SAN5} Santos N.~C., Israelian, G., Mayor, M., et al.~2005, A\&A 437, 1127



 

\bibitem{SMT} Smith, V.~V., Cunha, C., \& Lazzaro, D. 2001, AJ, 121, 3207

\bibitem{SND} Sneden, C.~1973, Ph.~D.~Thesis, University of Texas

%\bibitem{UMS}   Udry, S., Mayor, M., \& Santos N.~C. 2003, A\&A, 407, 369

%\bibitem{UM}   Udry, S., \& Mayor, M. 2001 in: Exo/Astrobiology. Proceedings of the first European Workshop (ESA Publication Division)

%\bibitem{UM2} Udry, S., \& Mayor, M. 2001 in: Astrobiology. Lecture Notes in Physics


\bibitem{lib1} Lewis, J., S.~1995 in: ``Physics and Chemistry of the Solar System'', Academic Press (San Diego)

\bibitem {} Th{\' e}venin, F. \& Idiart, T.~P. ~1999, ApJ, 521, 753-763

\bibitem{TAK} Takeda, Y., Sato, B., Kambe, E., et al. 2001, PASJ, 53, 1211
\bibitem{TH} Thielemann K-F., Argast, D., Brachwitz, F., et al. 2002, A\&SS, 281, 25
\bibitem{TIMM} Timmers, F.~X., Woosley, S.~E., \& Weaver, T.~A., 1995, ApJS, 98, 617

\bibitem{VAU} Vauclair, S. 2003, ApJ, 605, 874

\end{thebibliography}
\end{document}